\documentclass[a4paper,12pt]{article}
\usepackage[utf8]{inputenc}
\usepackage{geometry}
\usepackage[T1]{fontenc}
\usepackage{textcomp}
\usepackage{titlesec}

\usepackage{tabulary}
\usepackage{array}
\usepackage{makecell}
\usepackage{etoolbox}

\usepackage{hyperref}
\usepackage{natbib}
\usepackage{enumerate}
\usepackage{graphicx}
\usepackage{xcolor}
\usepackage{multirow}
\usepackage{textcomp,booktabs}
\usepackage{amssymb}
\usepackage{amsmath}
\usepackage{mathtools}
\usepackage{lscape}
\usepackage{rotating}
\usepackage{lineno}
\usepackage{bbm}
\usepackage{lmodern}

\usepackage{amsfonts}
\usepackage[ruled,vlined]{algorithm2e}  

\usepackage{animate}
\usepackage{adjustbox}

\usepackage{pdflscape}
\usepackage{afterpage}
\usepackage{capt-of}
\usepackage[section]{placeins}
\usepackage[title]{appendix}
\usepackage{caption}
\usepackage{subcaption}
\usepackage{chngcntr}
\usepackage{changepage}
\usepackage[flushleft]{threeparttable}
\usepackage{pdfpages}
\usepackage{mwe}
\usepackage{colortbl}
\usepackage{url}
\usepackage{authblk}
\usepackage{diagbox}
\usepackage{bm}
\usepackage{nameref}

\renewcommand{\baselinestretch}{1.5}

\title{A Bayesian Ensemble Projection of Climate Change and Technological Impacts on Future Crop Yields}

\author{
Dan Li\textsuperscript{1,2}\thanks{Email: d33.li@qut.edu.au}, 
Vassili Kitsios\textsuperscript{3}\thanks{Email: Vassili.Kitsios@csiro.au}, 
David Newth\textsuperscript{4}\thanks{Email: David.Newth@csiro.au}, 
Terence John O'Kane\textsuperscript{5}\thanks{Email: Terence.O'Kane@csiro.au}
}

\affil{\textsuperscript{1}School of Mathematical Sciences, Queensland University of Technology, Brisbane, Australia}
\affil{\textsuperscript{2}Environment, Commonwealth Scientific and Industrial Research Organisation, Dutton Park, QLD, Australia}
\affil{\textsuperscript{3}Environment, Commonwealth Scientific and Industrial Research Organisation, Aspendale, VIC, Australia}
\affil{\textsuperscript{4}Environment, Commonwealth Scientific and Industrial Research Organisation, Canberra, ACT, Australia}
\affil{\textsuperscript{5}Environment, Commonwealth Scientific and Industrial Research Organisation, Hobart, TAS, Australia}

\affil{\textsuperscript{*}Corresponding author}

\date{}

\begin{document}

\maketitle

\providecommand{\keywords}[1]
{
	\textbf{\textit{Keywords ---}} #1
}

\begin{abstract}

This paper introduces a Bayesian hierarchical modeling framework within a fully probabilistic setting for crop yield estimation, model selection, and uncertainty forecasting under multiple future greenhouse gas emission scenarios. By informing on regional agricultural impacts, this approach addresses broader risks to global food security. Extending an established multivariate econometric crop-yield model to incorporate country-specific error variances, the framework systematically relaxes restrictive homogeneity assumptions and enables transparent decomposition of predictive uncertainty into contributions from climate models, emission scenarios, and crop model parameters. In both in-sample and out-of-sample analyses focused on global wheat production, the results demonstrate significant improvements in calibration and probabilistic accuracy of yield projections. These advances provide policymakers and stakeholders with detailed, risk-sensitive information to support the development of more resilient and adaptive agricultural and climate strategies in response to escalating climate-related risks.

\end{abstract}



\section{Introduction}

Reliable quantification of uncertainty in crop-yield projections under climate change is critical for global food security. 
Beyond directly influencing food availability, crop production has broader socioeconomic and political implications. Fluctuations in agricultural outputs can trigger global price shocks, leading to social unrest and political instability, as exemplified by the Arab Spring, which was accelerated by climate-induced declines in wheat production and resulting price spikes (\citealp{johnstone2011global}). Additionally, agricultural forecasts serve as vital inputs to integrated assessment models (IAMs), shaping regional investment policies and adaptation strategies \citep{nelson2014climate}.
Accurate yield forecasts support effective adaptation strategies, mitigate risks to agricultural production, and guide policymaking in a rapidly changing climate (\citealp{rosenzweig2014assessing}; \citealp{asseng2013uncertainty}; \citealp{muller2021exploring}). Crop yields respond to numerous interacting factors, including climate extremes, biophysical processes, technological advancements, and socio-economic conditions. Together, these factors introduce significant complexity and uncertainty into yield forecasts (\citealp{challinor2018improving}; \citealp{muller2021exploring}). Addressing these uncertainties rigorously remains a fundamental challenge for climate-impact assessments and agricultural policy.

Earlier research has examined multiple approaches to reduce uncertainty in yield projections. A common method is to enhance crop-model representations of key physiological processes, such as crop sensitivity to temperature. These improvements have notably reduced simulation errors and uncertainty in crop-yield forecasts (\citealp{wang2017uncertainty}). \citet{zhang2015prediction} explicitly compared uncertainties from climate and crop models in maize yield predictions. They demonstrated that, over certain future periods, variability in climate projections often surpasses uncertainty arising from differences among crop models. More recently, studies have combined machine learning with global gridded crop models (GGCMs) and detailed environmental datasets. These hybrid methods successfully capture complex yield responses to extreme weather events, pests, and diseases. Consequently, they reduce uncertainty and reintroduce stressors often overlooked in earlier models (\citealp{li2023integrating}; \citealp{heinicke2022global}).

Despite these advances, most existing approaches remain focused primarily on refining model structures or incorporating additional data. Few studies have comprehensively characterized the full spectrum of uncertainty within a unified framework. Even fewer explicitly consider regional differences or systematically integrate uncertainties from multiple sources simultaneously. Ensemble-based studies (\citealp{asseng2013uncertainty}; \citealp{muller2021exploring}) highlight the value of multimodel comparisons, but typically assume uniform error distributions across regions. They also often rely on deterministic or semi-probabilistic methods, potentially underestimating or mischaracterizing the true scale of uncertainty (\citealp{franke2020ggcmi}). This leaves a significant gap: reliable and fully probabilistic quantification of yield uncertainty remains incomplete.

To fill this gap, our study introduces a Bayesian hierarchical model embedded in a fully probabilistic framework. Our approach enhances the precision and reliability of uncertainty quantification for crop-yield projections. We build upon the recent multivariate autoregressive crop-yield growth model (\citealp{li2025machine}) and introduce a novel extension that explicitly models error variances separately by country. This improvement relaxes the traditional assumption of homogeneous errors, better capturing real-world regional variability. Unlike previous approaches focused primarily on model refinement and data integration (\citealp{wang2017uncertainty}; \citealp{zhang2015prediction}), our framework fundamentally advances the quantification and interpretation of uncertainty itself. 
In our Bayesian framework, we jointly account for the uncertainties stemming from climate-model projections, emissions pathways, and the parameterisation of the crop model.
This approach provides detailed annual probabilistic forecasts and explicitly quantifies tail risks associated with extreme yield outcomes. Specifically, we measure extreme events using two established metrics from risk management: Value-at-Risk (VaR), the yield threshold that our model expects will be met or exceeded with $1-\alpha$ probability in that particular year, and Expected Shortfall (ES), the average yield conditional on falling below that VaR level. 
Because both numbers are estimated directly from the Bayesian posterior predictive distribution, they carry forward the full uncertainty from climate models, emissions pathways, and crop-model parameters, giving stakeholders a clear, year-by-year gauge of downside agricultural risk.

We validate the accuracy of our uncertainty quantification using historical crop and climate data. Through detailed global wheat production analyses, our model consistently outperforms existing methods. It provides robust, interpretable forecasts, offering policymakers clear, risk-informed guidance. This comprehensive framework thus enables more resilient adaptation planning and more informed agricultural decisions in the face of future climate extremes.

The remainder of this paper proceeds as follows. Section \ref{sec_data} describes the climate and crop datasets and explains our data processing methods. Section \ref{sec_methods} details our modeling framework. First, we review the existing multivariate autoregressive crop-yield growth model (MAR-X). Next, we describe its Bayesian implementation. Finally, we extend this model with a hierarchical structure to allow for country-specific error variances. Section \ref{sec_smc} presents the Sequential Monte Carlo (SMC) algorithm used for parameter estimation and model selection. Section \ref{sec_emp} provides our empirical results. Specifically, Section \ref{sec_history} evaluates historical forecast accuracy and tail risk. Section \ref{sec_project} shows probabilistic yield projections under various future climate scenarios. Section \ref{sec_discussion} concludes the paper by discussing key implications of our results.


\section{Data} \label{sec_data}
This study utilizes the same climate and crop datasets as described in \citet{li2025machine}, where a comprehensive description of the data sources and processing methods is provided. For completeness, we briefly summarize the relevant datasets below.

\subsection*{Climate data}

We use historical climate data from the Climate Research Unit (CRU) Time Series version 4.05 (CRU TS 4.05; \href{http://www.cru.uea.ac.uk}{http://www.cru.uea.ac.uk}), which provides monthly observations of surface climate variables, including temperature and precipitation, on a $0.5^\circ$ latitude–longitude grid from 1961 to 2018 \citep{harris2020version}. CRU TS is a widely used observational dataset that is constructed from weather station records using interpolation methods. It belongs to a broader category of climate reanalysis and observational reconstruction products, which aim to reconstruct past climate conditions by integrating diverse observational datasets \citep{kalnay2018ncep, dee2011era}.

To generate future climate projections, we apply the QuickClim method proposed by \citet{kitsios2023machine}. QuickClim is a machine-learning-based emulator designed to efficiently replicate the outputs of complex global climate models (GCMs). A climate model is a computer-based simulation that represents how different parts of the Earth's climate system—such as the atmosphere, oceans, land surface, and ice—interact over time. While GCMs provide detailed and scientifically robust simulations of climate responses under various greenhouse gas and land-use scenarios, they are computationally intensive. QuickClim addresses this limitation by rapidly approximating monthly climate variables, including temperature and precipitation, across a wide range of emissions scenarios.

To align with standard climate-impact assessments, we focus on three Representative Concentration Pathways (RCPs)—RCP 2.6, RCP 4.5, and RCP 8.5—corresponding to low, intermediate, and high greenhouse-gas emissions. These pathways were adopted by the Coupled Model Intercomparison Project Phase 5 (CMIP5) as common inputs for global climate-model experiments \citep{van2011representative}. CMIP5 itself is an internationally coordinated ensemble of climate models that simulates future climate conditions under harmonized assumptions about greenhouse-gas concentrations and land use \citep{taylor2012overview}. Each RCP traces a plausible evolution of emissions and land use that delivers a specified level of radiative forcing by 2100. Radiative forcing measures the net energy imbalance introduced into the Earth system by elevated greenhouse-gas levels and other agents, and it can be translated into effective atmospheric concentrations of carbon-dioxide equivalents ($CO_{2e}$). Figure \ref{fig:Combined_Figure_QC}a plots those $CO_{2e}$ trajectories for the RCPs alongside the ensemble of 100 QuickClim pathways used in our study.

\subsection*{Crop data}

Annual historical crop production data spanning the years 1961–2018 are obtained from the FAOSTAT database (\url{http://faostat.fao.org}), covering national-level wheat production. Information about crop planting and harvesting dates, necessary for linking climate variables with crop growing seasons, is taken from the global crop calendar database of \citet{sacks2010crop}. Using these growing periods, we extract relevant climate variables during the crop growth stages. In cases where countries have multiple cropping seasons, we average seasonal climate conditions within each country before associating them with annual crop production records. Additionally, we apply a standard filtering process, removing outliers identified based on anomalously high volatility in annual crop production. This ensures robust estimates by minimizing potential biases arising from data inconsistencies or extreme events. While this work focuses on wheat in the empirical analysis, the proposed approach can be applied to other crops. 

\section{Methods} \label{sec_methods}
\subsection{MAR-X Crop Yield Growth Model} 

The recently developed multivariate autoregressive model with exogenous climate variables for the estimation of crop yield log returns, (MAR-X; \citealp{li2025machine}),
takes the form of:
\begin{align}
\mathbf{y}_t &= \theta_0(t) + \theta_1 \mathbf{y}_{t-1} + \theta_2 \mathbf{y}_{t-2} + \theta_3 \Delta \mathbf{T}_{t} + \theta_4 \Delta \mathbf{T}_{t}^2 + \bm{\epsilon}_t, \label{model_TVI_2} \\
\theta_0(t) &= a \cdot e^{-\lambda t}. \label{model_TVI_intercept}
\end{align}
Here $\mathbf{y}_t = [y_{1,t}, y_{2,t}, \dots, y_{K,t}]'$ denotes the log-returns of wheat yield for $K$ crop-growing regions at year $t$. The parameters $\theta_1, \theta_2, \theta_3, \theta_4$ are assumed to be scalars, $\bm{\epsilon}_t$ is the error term, and $\Delta \mathbf{T}_t$ and $\Delta \mathbf{T}_t^2$ represent the differences in the annual mean temperature during the crop-growing period and its quadratic form, respectively, for the current year $t$ and the previous year $t-1$ across $K$ regions. 
Previous work \citep{li2025machine} evaluated multiple functional forms for the intercept term, $\theta_0(t)$, and found that an exponentially decaying form (Eq \ref{model_TVI_2}) provided the best fit to historical data and improved forecasting performance for wheat.

While the MAR-X provides valuable insights into estimating wheat yield log returns, it has several limitations that highlight areas for potential improvement. 
The model assumes a shared error term across all countries, which may not accurately represent the diverse sources of uncertainty in agricultural production. The shared error term overlooks country-specific factors that could influence yield variability, such as localized climate anomalies, policy changes, or infrastructure variations. This limitation potentially reduces the precision and robustness of yield predictions across different countries. 
Additionally, the model does not consider uncertainties inherent to the agricultural production model itself. This gap suggests a need for a more comprehensive approach that rigorously quantifies and incorporates these uncertainties to enhance the reliability of yield and production projections.


\subsection{B-MAR-X: A Bayesian Framework for MAR-X} \label{MAR-X_current}
The Bayesian framework requires the specification of prior distributions for each model parameter. For the intercept parameters, $a$ and $\lambda$, we assume uniform priors bounded within specific ranges. The parameters \(a\) and \(\lambda\) in Eq \ref{model_TVI_intercept} are sampled as:
\begin{equation}
    a \sim \text{Uniform}(0, 0.5), \quad \lambda \sim \text{Uniform}(0, 0.2).
\end{equation}
The above prior ensures that both the initial technological advancements and the decay rates remain positive. The upper bounds are selected to include the non-linear least squares point estimates from the non-Bayesian MAR-X model, while maintaining a narrow range to enhance the efficiency of the Bayesian sampling process.


The autoregressive coefficients, $\theta_1$ and $\theta_2$ (Eq \ref{model_TVI_2}), which relate to the lagged yield log-returns, are assigned normal priors with mean 0 and standard deviation 0.5:
\begin{equation}
    \theta_1 \sim \mathcal{N}(0, 0.5), \quad \theta_2 \sim \mathcal{N}(0, 0.5).
\end{equation}
The normal distributions of the above priors reflect the possibility of both positive and negative effects, and a small variance is selected to be consistent with the relatively modest point estimates reported in the existing literature.
The temperature-related coefficients, $\theta_3$ and $\theta_4$, are also modeled with normal priors, assuming mean zero and and standard deviations reflecting more conservative beliefs about their influence:
\begin{equation}
    \theta_3 \sim \mathcal{N}(0, 0.2), \quad \theta_4 \sim \mathcal{N}(0, 0.1).
\end{equation}
Finally, the variance of the error term, $\sigma^2$, is assigned a commonly used non-informative inverse-gamma prior, which is a natural and appropriate choice for positive variance parameters:
\begin{equation}
    \sigma^2 \sim \text{Inverse-Gamma}(2, 1).
\end{equation}
The posterior distributions of the model parameters are obtained by combining the specified priors with the likelihood of the observed data. 

The model specification for each region \(i\) at time \(t\) is given by
\begin{equation}
y_{i,t} = \theta_0(t) + \theta_1 y_{i,t-1} + \theta_2 y_{i,t-2} + \theta_3 \Delta T_{i,t} + \theta_4 \Delta T_{i,t}^2 + \epsilon_{i,t},
\end{equation}
where the conditional mean is defined as
\begin{equation}
\mu_{i,t} = \theta_0(t) + \theta_1 y_{i,t-1} + \theta_2 y_{i,t-2} + \theta_3 \Delta T_{i,t} + \theta_4 \Delta T_{i,t}^2. 
\label{MAR-X-Mean}
\end{equation} 
The likelihood is based on the assumption that the error terms are independently and identically distributed as $\epsilon_{i,t} \sim \mathcal{N}(0,\sigma^2)$, for $i=1,\dots,K$ regions and $t=1,\dots,T$ time periods.
The density for each observation is then given by
\begin{equation}
p(y_{i,t} \mid \Theta) = \frac{1}{\sqrt{2\pi}\sigma} \exp\!\left\{-\frac{1}{2\sigma^2}\left(y_{i,t} - \mu_{i,t}\right)^2\right\},
\end{equation}
with \(\Theta = \{a, \lambda, \theta_1, \theta_2, \theta_3, \theta_4, \sigma^2\}\) representing the model parameters.

Assuming that the observations are conditionally independent across regions and time, the likelihood for the full dataset is
\begin{equation}
L(\Theta) = \prod_{t=1}^{T} \prod_{i=1}^{K} \frac{1}{\sqrt{2\pi}\sigma} \exp\!\left\{-\frac{1}{2\sigma^2}\left(y_{i,t} - \mu_{i,t}\right)^2\right\}.
\end{equation}
The Bayesian framework for the MAR-X model (B-MAR-X) provides a flexible and probabilistic approach to estimating wheat yield log-returns while accounting for parameter uncertainty and variability.

\subsection{HB-MAR-X: A Hierarchical Bayesian Approach to Crop Yield Modelling} \label{BHM}
In order to address the limitations inherent in the single, pooled error specification of the MAR-X, we now introduce a more sophisticated framework. In this section, we extend our analysis by incorporating a hierarchical structure that assigns country‑specific error distributions while still borrowing strength across the panel. 
By nesting the individual variances in a common hyper‑prior, the model simultaneously (i) captures genuine cross‑country heterogeneity in yield volatility, (ii) mitigates over‑fitting in data‑sparse settings through partial pooling, and (iii) delivers fully probabilistic forecasts that propagate both parameter and process uncertainty. 
This hierarchical structure improves information sharing across countries and yields more precise, well‑calibrated uncertainty estimates—providing a firmer basis for data‑driven food‑security risk assessments. Details of the model formulation and prior specifications are provided below.



The  conditional mean of the crop yield log-returns for $K$ countries at time $t$ follows the MAR-X model specification given in Eq \ref{model_TVI_2}, where $\bm{\epsilon}_t = [\epsilon_{1,t}, \epsilon_{2,t}, \dots, \epsilon_{K,t}]'$ (with $'$ denoting the transpose) is a vector of $i.i.d.$ error terms for each country. 
Unlike the standard MAR-X model, here we assume that the error term for each country follows its own normal distribution: 
\begin{equation}
\epsilon_{i,t} \sim \mathcal{N}(0, \sigma_i^2), \quad \text{for } i = 1, 2, \dots, K,
\end{equation}
where $\sigma_i^2$ represents the variance of the error term for country $i$. Rather than fixing the variance $\sigma_i^2$, we model them as random effects drawn from inverse-gamma distributions. Specifically, the hierarchical structure for the error variances is given by:
\begin{equation}
    \sigma_i^2 \sim \text{Inverse-Gamma}(\alpha_\sigma, \beta_\sigma), \quad \text{for } i = 1, 2, \dots, K.
\end{equation}\label{HB_MAR-X}
In this structure, $\alpha_\sigma$ is a global shape parameter that determines the general behavior of error variability across countries, while $\beta_\sigma$ is a global scale parameter. The hyperpriors for $\alpha_\sigma$ and $\beta_\sigma$ are defined as:
\begin{equation}
    \alpha_\sigma \sim \text{Gamma}(2, 1), \ \beta_\sigma \sim \text{Gamma}(2, 1).
\end{equation}
This hierarchical setup allows for flexible modeling of country-specific error variances, with global priors capturing the overall distribution of variability.


Next, we present the likelihood function for the hierarchical Bayesian MAR‑X (HB-MAR-X) model.  
Let $\mathbf{Y}=\{y_{i,t}\}_{i=1,\dots ,K;\,t=3,\dots ,T}$ denote the panel of observed crop‑yield log‑returns, and let the conditional mean $\mu_{i,t}$ be defined as in Eq \ref{MAR-X-Mean}.
For each time point $t$, define $\boldsymbol\epsilon_t=[\epsilon_{1,t},\dots ,\epsilon_{K,t}]'$ with  
$\epsilon_{i,t}=y_{i,t}-\mu_{i,t}$.  
Assuming mutually independent Gaussian innovations,
\begin{equation}
  \boldsymbol\epsilon_t\mid\boldsymbol\sigma^{2}
  \;\sim\;
  \mathcal N\!\bigl(\mathbf 0,\operatorname{diag}(\sigma_1^{2},\dots ,\sigma_K^{2})\bigr),
  \qquad t=3,\dots ,T.
\end{equation}
The likelihood contribution for each time point $t$, conditional on the structural coefficients  
$\boldsymbol{\theta}=(a,\lambda,\theta_{1},\theta_{2},\theta_{3},\theta_{4})$  
and the vector of country‑level variances  
$\boldsymbol{\sigma}^{2}=(\sigma_{1}^{2},\dots ,\sigma_{K}^{2})$, is given by: 
\begin{equation}
  P(\mathbf{y}_t \mid \boldsymbol{\theta},\boldsymbol{\sigma}^{2})
  = (2\pi)^{-K/2}
    \Bigl(\prod_{i=1}^{K}\sigma_i^{-1}\Bigr)
    \exp\!\Bigl\{-\frac{1}{2}
      \sum_{i=1}^{K}\frac{\bigl(y_{i,t}-\mu_{i,t}\bigr)^{2}}{\sigma_i^{2}}\Bigr\}.
\end{equation}
Assuming conditional independence across time, the likelihood of the full sample is
\begin{equation}
  \mathcal{L}(\boldsymbol{\theta},\boldsymbol{\sigma}^{2};\mathbf{Y})
  =\prod_{t=3}^{T}
     P(\mathbf{y}_t \mid \boldsymbol{\theta},\boldsymbol{\sigma}^{2}).
\end{equation}
This form first groups the independent Gaussian densities by time point and then multiplies across $t$.  
It highlights two features of the hierarchical MAR‑X specification:  
(i) conditional independence across countries within each time slice, and 
(ii) heteroscedasticity introduced by the country‑specific variances $\sigma_i^{2}$, which are themselves random quantities in the next layer of the hierarchy.

The HB-MAR-X model retains the original econometric framework’s simplicity while introducing the flexibility of country‑specific random effects. Independent Gaussian shocks remain at the observation level. However, their variances are now modeled using a shared Inverse‑Gamma distribution, with hyper‑parameters estimated from the data. This structure aims to enhance the adaptability and robustness of the model by accounting for both global tendencies and local variations in crop yield responses. 
This layered design yields several practical benefits: it disciplines extreme variance estimates, improves out‑of‑sample predictive accuracy, and furnishes decision‑makers with posterior predictive distributions that fully reflect ensemble uncertainties across climate scenarios and technological pathways. In short, the model offers a robust, policy‑relevant platform for quantifying both global trends and local idiosyncrasies in future crop‑yield dynamics.

\FloatBarrier

\section{Model Estimation and Selection}
\label{sec_smc}

The Bayesian estimation procedure adopted in this study utilizes Sequential Monte Carlo (SMC), a powerful approach suited for hierarchical models where posterior distributions may be complex or multimodal. SMC efficiently approximates posterior distributions and computes marginal likelihoods, facilitating robust uncertainty quantification and model comparison. The detailed SMC algorithm used for parameter estimation and model selection is described fully in \ref{SMC}. 

To provide a visual illustration of how the uncertainties surrounding the crop yield model's unknown parameters are quantified and distinguished from point estimates, Figure \ref{fig:posterior_marx} presents the marginal posterior distributions for the B-MAR-X model described in section \ref{MAR-X_current}, alongside point estimates generated by the existing frequentist approach in \citet{li2025machine}. 
While the frequentist method produces point estimates without explicitly incorporating uncertainty in the error variance, the Bayesian approach jointly models both the parameters and the error variance, resulting in a full probabilistic characterization of parameter uncertainty. The posterior distribution of the error variance is also included in the figure to illustrate this aspect. 

For all parameters of the B-MAR-X model, the non-linear least squares point estimates fall well within the corresponding Bayesian posterior distributions and align with their high-density regions. This consistency between the two approaches supports the validity of the Bayesian estimates and highlights the effectiveness of the uncertainty quantification achieved through the SMC-based inference. 
For the HB-MAR-X model, instead of using a single variance parameter as in the B-MAR-X model, the posterior distributions of country-level variances and the associated joint parameters are estimated jointly.
The accuracy of the uncertainty estimates is further evaluated in the results section.

In addition to parameter estimation, Bayesian methods facilitate rigorous model selection through the computation of marginal likelihoods (model evidence). For the hierarchical Bayesian model introduced in section \ref{BHM}, the marginal likelihood involves integrating over both model parameters and hyperparameters. A complete derivation and description of this marginal likelihood, including its estimation via SMC methods, is provided in \ref{sec:model_selection}.

Specifically, for the proposed hierarchical model, the marginal likelihood \( P(\mathbf{y}) \) of the observed data is defined as:
\begin{equation}
    P(\mathbf{y}) = \iint P(\mathbf{y} \mid \bm{\Theta}) \, P(\bm{\Theta} \mid \bm{\Phi}) \, P(\bm{\Phi}) \, \mathrm{d}\bm{\Theta} \, \mathrm{d}\bm{\Phi},
\end{equation}
where $\bm{\Theta} = \left\{ a, \lambda, \theta_1, \theta_2, \theta_3, \theta_4, \sigma_i^2 \mid i = 1, \dots, K \right\}$ denotes the model parameters and their country level variances, and $\bm{\Phi} = \left\{ \alpha_\sigma, \beta_\sigma \right\}$ denotes the hyperparameters.
The likelihood of the observed data \( \mathbf{y} = \{\mathbf{y}_1, \mathbf{y}_2, \dots, \mathbf{y}_T\} \), given the model parameters \( \bm{\Theta} \), is:
\begin{equation}
    P(\mathbf{y} \mid \bm{\Theta}) = \prod_{t=2}^T P(\mathbf{y}_t \mid \mathbf{y}_{t-1}, \mathbf{y}_{t-2}, a, \lambda, \theta_1, \theta_2, \theta_3, \theta_4, \sigma_i^2),
\end{equation}
where \( P(\mathbf{y}_t \mid \cdot) \) follows a Gaussian distribution specified in section \ref{BHM}.
The prior distribution of the model parameters, given the hyperparameters \( \bm{\Phi} \), is:
\begin{equation}
    P(\bm{\Theta} \mid \bm{\Phi}) = \prod_{i=1}^K P(a, \lambda, \theta_1, \theta_2, \theta_3, \theta_4, \sigma_i^2 \mid \alpha_\sigma, \beta_\sigma).
\end{equation}
The prior distribution of the hyperparameters \( \bm{\Phi} \) is:
\begin{equation}
    P(\bm{\Phi}) = P(\alpha_\sigma) \, P(\beta_\sigma).
\end{equation}
The marginal likelihood \( P(\mathbf{y}) \) for this hierarchical model can be efficiently estimated using SMC by incrementally computing the normalizing constants along an annealed sequence of posterior distributions, thereby providing a practical method for quantifying model evidence and performing Bayesian model comparison.

To construct a robust and parsimonious multivariate crop-yield model, we employ a forward selection procedure based on Bayesian model evidence. Initially, the model includes the 15 largest wheat-producing countries, ensuring a solid foundation of informative data. Subsequently, additional countries are sequentially incorporated one at a time. For each newly added country, we compute the corresponding marginal likelihood (model evidence) using SMC as described above. 
If the inclusion of a country leads to an increase in the overall model evidence, which indicates improved predictive accuracy relative to the complexity introduced, it is retained in the model. Conversely, if the addition of a country results in a reduction in model evidence, suggesting an adverse impact due to increased complexity or noise, it is excluded from the final specification. 
This approach strategically departs from existing modeling practices, where typically all available countries are simultaneously incorporated, potentially introducing unnecessary noise and complexity. In contrast to traditional methodologies employing non-linear least squares point estimators, our proposed selection procedure leverages Bayesian model evidence computed through SMC, thereby providing substantial methodological advantages by inherently accounting for parameter uncertainty and effectively penalizing model complexity.

\begin{table}[htbp]
\centering
\begin{tabular}{|>{\centering\arraybackslash}p{6.5cm}|>{\centering\arraybackslash}p{4.0cm}|>{\centering\arraybackslash}p{5.0cm}|}
\hline
 & \multicolumn{2}{c|}{\textbf{Estimation of Log Evidence}} \\
\cline{2-3}
\textbf{Models} & {\textbf{All 57 Countries}} & {\textbf{Selected 40 Countries}} \\
\hline
B-MAR-X & 663 & 1189 \\
\hline
HB-MAR-X ($a$, $\lambda$) & 660 & 1185\\
\hline
HB-MAR-X ($\sigma^2$) & $\bm{1266}$ & $\bm{1408}$ \\
\hline
HB-MAR-X ($a$, $\lambda$, $\sigma^2$) & ${1192}$ & ${1389}$ \\
\hline
MAR-X-IND ($a$, $\lambda$, $\sigma^2$) & -51 & 180 \\

\hline
HB-MAR-X (all model parameters) & -755 & 143\\
\hline
\end{tabular}
\caption{Estimation of log evidence for different models using SMC.}
\label{tab:log_evidence}
\end{table}

By utilizing the SMC method for computationally efficient estimation of marginal likelihoods in this complex, high-dimensional framework, our iterative Bayesian approach ensures that only those countries substantially enhancing explanatory power and predictive accuracy are retained. Consequently, our forward-selection procedure systematically optimizes model interpretability, efficiency, and forecasting reliability, surpassing traditional least-squares-based techniques through a principled balance between model parsimony and comprehensive predictive capability.

Table \ref{tab:log_evidence} presents the model selection results based on the estimated log model evidence for several candidate specifications. The full list of wheat-producing countries considered in the literature is provided in Table \ref{crop_countries}, with the countries excluded from the final model in this study indicated in bold and shaded gray. 
In addition to evaluating the proposed HB-MAR-X model, which incorporates a hierarchical structure for country-level error terms only, we also examine several alternative specifications. These include: (i) a model with hierarchical structures applied solely to the country-level time-varying intercepts, specifically the parameters $a$ and $\lambda$; (ii) a model incorporating hierarchical structures for both the country-level intercepts and error terms; and (iii) a fully hierarchical specification that includes hierarchical structures for all model parameters and error terms. These three variants are denoted as HB-MAR-X ($a$, $\lambda$), HB-MAR-X ($a$, $\lambda$, $\sigma^2$), and HB-MAR-X (all model parameters), respectively, as shown in Table \ref{tab:log_evidence}.  
Moreover, as an alternative to employing hierarchical structures with hyperpriors in the HB-MAR-X ($a$, $\lambda$, $\sigma^2$) specification, we also consider a model in which priors are assigned directly to the country-specific time-varying intercepts and error terms without a hierarchical structure. This model is referred to as MAR-X-IND ($a$, $\lambda$, $\sigma^2$).

The log evidence in Table \ref{tab:log_evidence} considers both the full set of 57 wheat-producing countries and a selected subset comprising 40 countries determined via our forward selection procedure. 
Across all model variants, there is a consistent improvement in model evidence when reducing the number of included countries from 57 to the selected subset of 40. This result highlights the benefit of selecting a subset of countries, demonstrating that removing countries that contribute primarily noise or irrelevant information significantly enhances model parsimony and predictive performance. 
Specifically, the HB-MAR-X ($\sigma^2$) model, which applies a hierarchical structure exclusively to country-level error terms, achieves the highest log evidence in both scenarios. This indicates that introducing country-specific variance structures substantially improves the model's capacity to capture heterogeneity in crop-yield data without excessive complexity. 
Simpler models such as the traditional B-MAR-X and the hierarchical model without variance terms (HB-MAR-X ($a, \lambda$)) achieve notably lower model evidence, highlighting the importance of modeling country-specific variances explicitly. 
Finally, the fully hierarchical model (HB-MAR-X with all parameters hierarchical) and the independent prior model (MAR-X-IND ($a, \lambda, \sigma^2$)) yield significantly poorer results, particularly when using all 57 countries. This indicates that excessive hierarchical complexity or independent priors without hierarchical constraints may lead to overfitting and reduce overall model performance.


\section{Crop Yield Projections and Risk Measures} \label{sec_emp}
\subsection{Analysis of Historical Forecasting Results and Tail Risk Measures} \label{sec_history}

To assess the suitability of an agricultural production model, for instance, one might consider its predictive capability. 
Building upon the Bayesian framework and SMC approach previously introduced, the posterior distribution \(\pi_{t}(\bm{\theta}|\bm{y}_{1:t})\) facilitates one-step-ahead posterior predictions for data at time \( t+1 \).  Within this fully Bayesian setting, the posterior predictive distribution for upcoming observations $\hat{\bm{y}}_{t+1}$ is obtained by integrating over the model parameters, thereby naturally incorporating parameter uncertainty without asymptotic approximations (see detailed derivation in \ref{appendix:forecasting}). 
The posterior predictive distributions can be approximated using the previously described SMC algorithm, generating weighted samples from the posterior. Point estimates, such as mean or median, and predictive intervals can then be straightforwardly computed from this weighted set.

\begin{figure}[htbp!]
  \centering
  \hspace{-2cm}
  \begin{subfigure}[b]{0.9\textwidth}
    \centering
    \includegraphics[height=7cm]{ 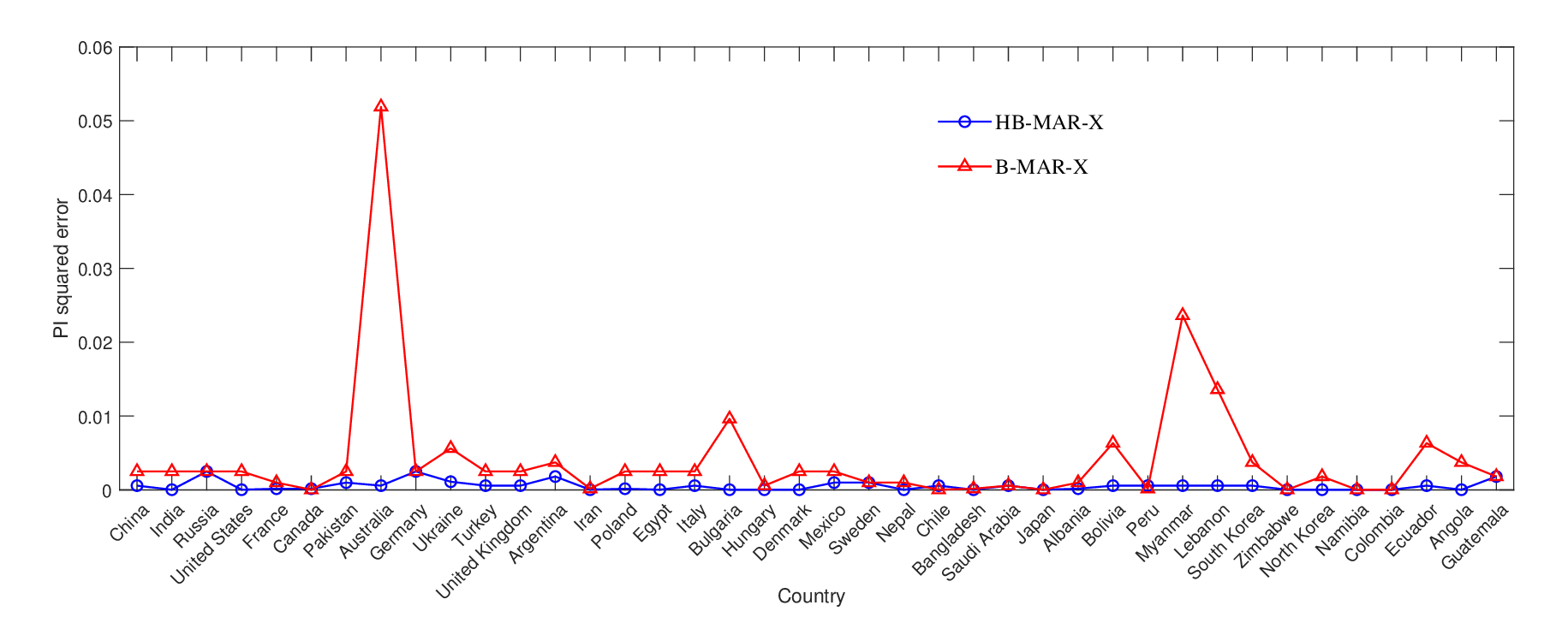}
    \caption{In-Sample}
    \label{fig:forecast_error_InSample}
  \end{subfigure}

  \hspace{-0.25cm} 
  \begin{subfigure}[b]{0.95\textwidth}
    \centering
    \includegraphics[height=7cm]{ 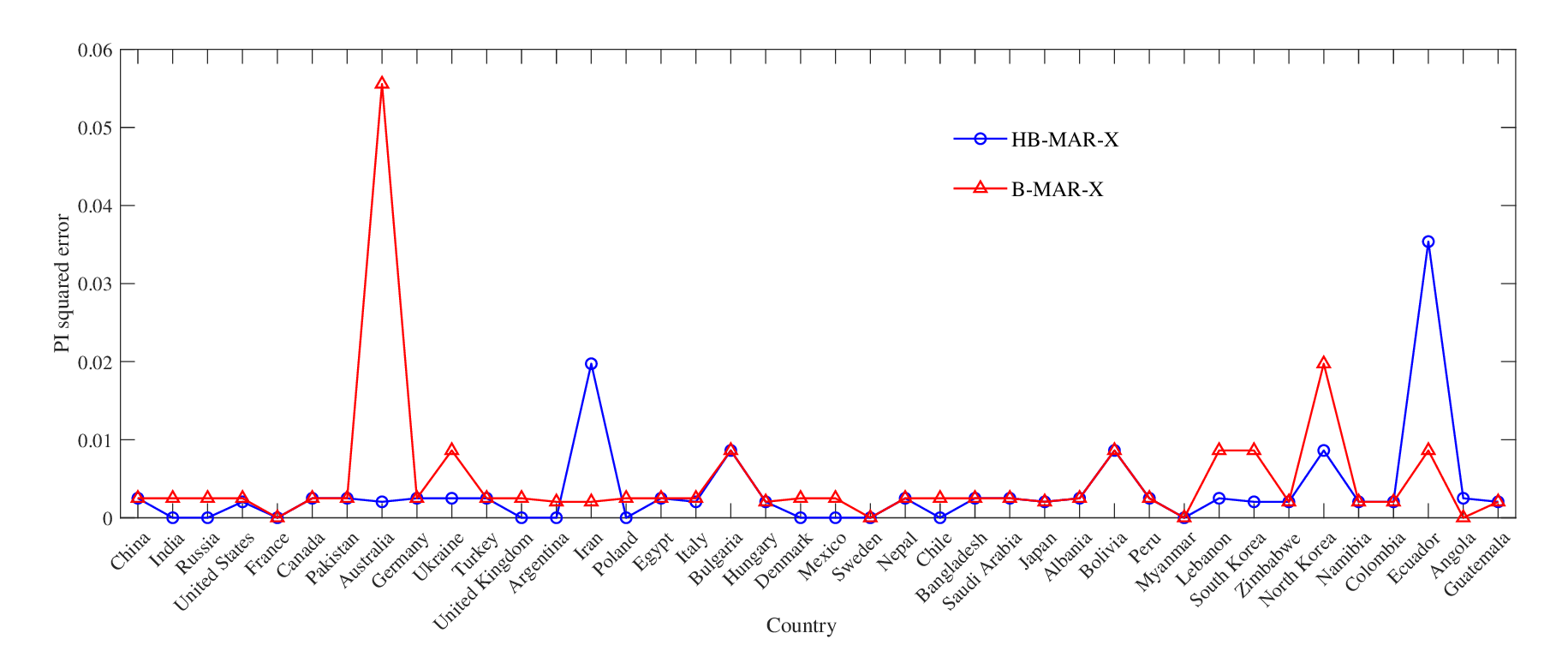}
    \caption{Out-of-Sample}
    \label{fig:forecast_error_OutSample}
  \end{subfigure}

  \caption{Squared Error of the 95\% Prediction Interval. 
  (a) In-sample results over 54 years, showing the squared difference between the observed frequency of values captured by the 95\% prediction interval and the expected 95\% coverage. 
  (b) Out-of-sample results over 21 years, similarly showing the deviation from ideal predictive coverage.}
  \label{fig:forecast_error_combined}
\end{figure}




To assess the quality of the predictive intervals produced by the model, we evaluate the accuracy of the posterior predictive distributions using the squared error of the 95\% prediction interval (PI). Specifically, for each country, we compute the squared difference between the empirical coverage rate—the proportion of observed data points falling within the 95\% predictive interval—and the nominal coverage rate of 0.95. This provides a direct measure of how well-calibrated the predictive intervals are, with smaller squared errors indicating better alignment between the forecast distribution and the realized outcomes.

Figure \ref{fig:forecast_error_combined} presents the results of this calibration test for two models: B-MAR-X (red triangles) and HB-MAR-X (blue circles). Panel (a) displays the in-sample performance over 54 years, while panel (b) shows the out-of-sample results based on a 21-year evaluation period excluded from model estimation. 
The in-sample results demonstrate that the HB-MAR-X model generally achieves lower squared errors across most countries, suggesting more accurate and reliable predictive intervals. In contrast, the B-MAR-X model exhibits pronounced miscalibration in certain countries, such as Australia and Myanmar, where the squared error spikes significantly, indicating worse interval coverage. These discrepancies likely arise from unmodeled heterogeneity or omitted hierarchical structure, which the HB-MAR-X model is explicitly designed to address.
The out-of-sample results generally support the in-sample findings. As shown in Figure \ref{fig:forecast_error_combined}(b), the HB-MAR-X model achieves lower or comparable squared errors relative to B-MAR-X across many countries, with notable improvements in a few cases. In particular, HB-MAR-X exhibits a dramatic reduction in squared error for Australia, where B-MAR-X clearly miscalibrates. While the differences are more modest for most other countries, the HB-MAR-X model maintains consistently low and stable squared errors, suggesting that its hierarchical structure contributes to more reliable generalization and improved uncertainty quantification.


We also rigorously evaluate the forecasting performance using leave-future-out cross-validation (LFO-CV) as proposed by \citet{burkner2020approximate}. Unlike traditional leave-one-out methods, LFO-CV explicitly considers temporal dependence, making it appropriate for sequentially ordered crop yield data. Predictive accuracy of the one-step-ahead forecasts is quantified by the expected log predictive density (elpd; \citealp{vehtari2017practical}), comprehensively evaluating the entire predictive distribution rather than single-point predictions (details of this evaluation are provided in \ref{appendix:forecasting}). 
The proposed HB-MAR-X ($\sigma^2$) model, which exhibits the highest log evidence as shown in Table \ref{tab:log_evidence}, achieved a higher elpd of 8.69 compared to 7.02 for the existing B-MAR-X model, indicating superior predictive performance. This improvement suggests that incorporating hierarchical variance components enhances the model's ability to accurately forecast crop yields across the selected 40 countries.


In our Bayesian framework for crop yield forecasting, the predictive distributions provide a foundation for assessing extreme agricultural risks. Borrowing concepts from financial risk management, we adapt the Value-at-Risk (VaR) and Expected Shortfall (ES) metrics to quantify the potential for adverse outcomes in agricultural production. Specifically, we use these measures both in terms of negative log yield returns (capturing relative yield losses) and in terms of low absolute yield levels, thereby making them intuitive and relevant to practical agricultural risk management.
Let \(Y\) denote a random variable representing yield-related risk. This may correspond to the negative log yield return, where higher values indicate greater yield loss, or it may represent the actual yield level, where lower values correspond to poorer production outcomes. Let \(\alpha \in (0,1)\) denote a chosen confidence level (commonly, \(\alpha = 0.0\) or \(\alpha = 0.99\)).

\begin{figure}[htbp!]
  \centering
  \captionsetup[subfigure]{skip=2pt}

  \begin{subfigure}[b]{1\textwidth}
    \centering
    \includegraphics[width=1\textwidth,height=7.5cm]{ 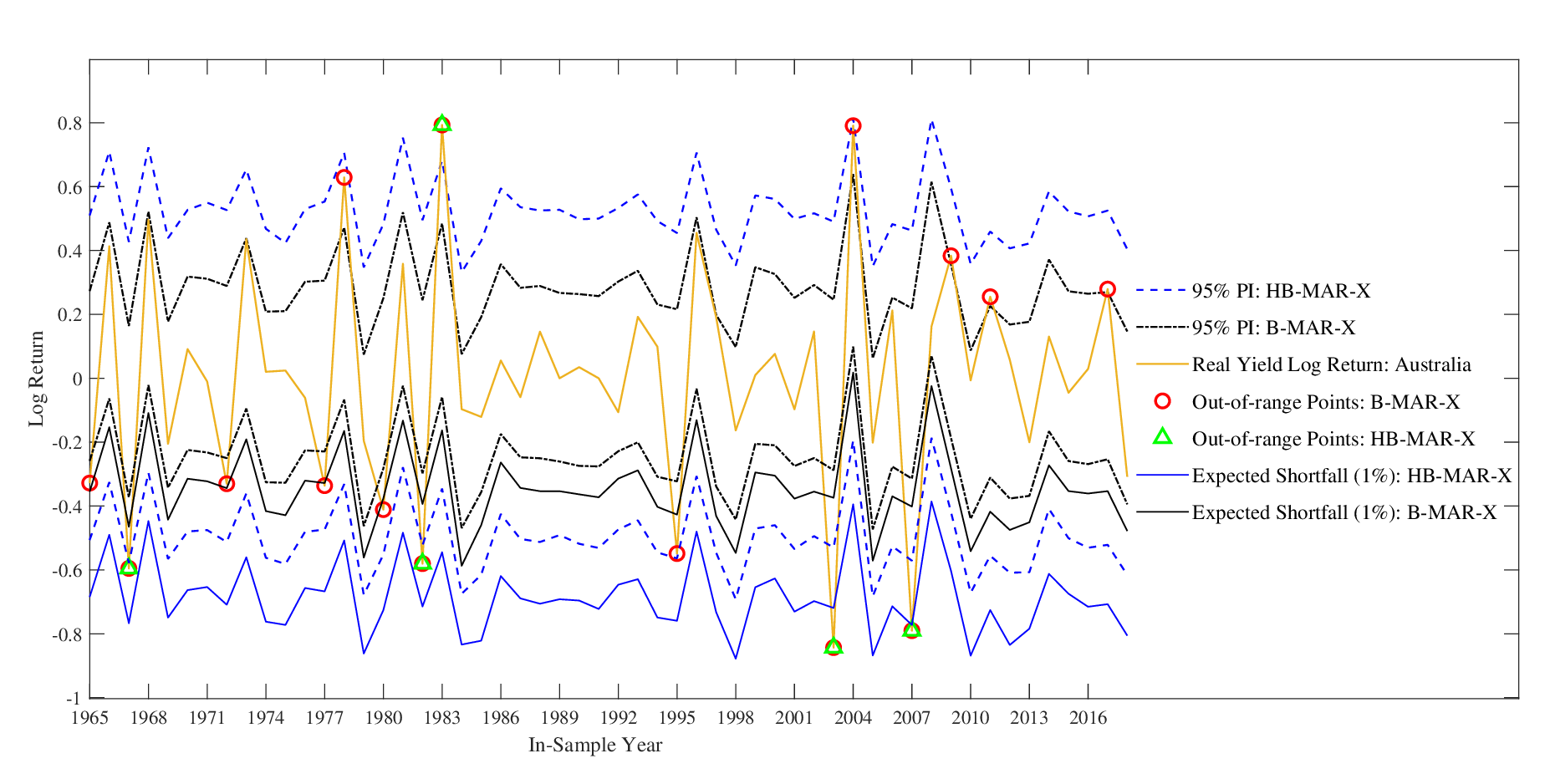}
    \caption{}
    \label{fig:forecast_Australia_InSample}
  \end{subfigure}


  \begin{subfigure}[b]{1\textwidth}
    \centering
    \includegraphics[width=1\textwidth,height=7.5cm]{ 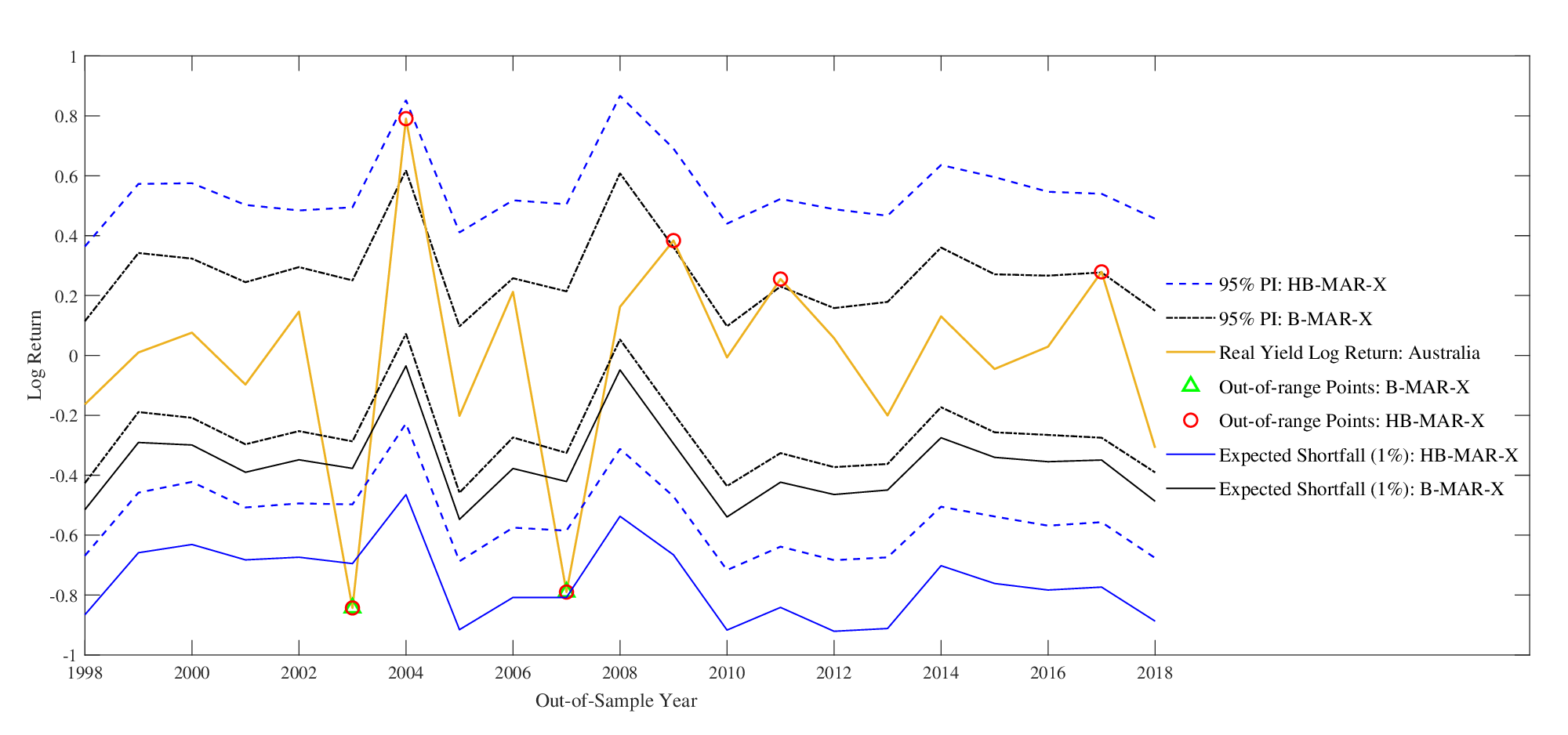}
    \caption{}
    \label{fig:forecast_Australia_OutSample}
  \end{subfigure}

  \caption{(a) In-sample 95\% prediction interval for the wheat yield log return in Australia over all 54 years. 
  (b) Out-of-sample 95\% prediction interval for the wheat yield log return in Australia over the 21 evaluation years.}
  \label{fig:forecast_Australia_combined}
\end{figure}

Value-at-Risk at confidence level \(\alpha\), denoted by \(\mathrm{VaR}_\alpha(Y)\), is defined as the \(\alpha\)-quantile of the distribution of \(Y\):
\begin{align}
    \mathrm{VaR}_\alpha(Y) = \inf \{ x \in \mathbb{R} \mid P(Y \le x) \ge \alpha \}.
\end{align}
Here, \(x\) represents a threshold for the yield-related quantity \(Y\), such that there is at least an \(\alpha\) probability that \(Y\) does not exceed \(x\).  When \(Y\) is the negative log yield return, \(x\) corresponds to a worst-case loss level not exceeded with high probability. When \(Y\) is the absolute production level, \(x\) identifies a minimum production level that is met or exceeded with probability \(\alpha\).

Expected Shortfall at level \(\alpha\), denoted by \(\mathrm{ES}_\alpha(Y)\) (also known as Conditional Value-at-Risk or CVaR), measures the average severity of outcomes in the tail beyond the VaR threshold. Assuming sufficient continuity of the distribution of \(Y\), ES is defined as:
\begin{align}
\mathrm{ES}_\alpha(Y) = \mathbb{E}\left[ Y \,\middle|\, Y \ge \mathrm{VaR}_\alpha(Y) \right],
\end{align}
where \(\mathbb{E}[\cdot]\) denotes the expectation operator. This formulation is appropriate when higher values of \(Y\) correspond to worse outcomes, such as in the case of negative log yield returns. In this context, the Expected Shortfall represents the average of the worst \(1-\alpha\) proportion of outcomes, capturing the expected magnitude of extreme yield losses. 
When instead \(Y\) represents a quantity where lower values are associated with poorer outcomes, such as absolute production levels, the definition is typically adapted to focus on:
\begin{align}
\mathrm{ES}_\alpha(Y) = \mathbb{E}\left[ Y \,\middle|\, Y \le \mathrm{VaR}_\alpha(Y) \right],
\end{align}
In this case, the Expected Shortfall captures the average production among the lowest-performing outcomes that occur with probability at least \(1-\alpha\), and thus quantifies the expected severity of low-production scenarios. We denote this as the $(1-\alpha)*100\%$ ES (for example, the 1\% ES).

\begin{figure}[htbp!]
  \centering
  \captionsetup[subfigure]{skip=2pt}

  \begin{subfigure}[b]{1\textwidth}
    \centering
    \includegraphics[width=1\textwidth,height=8cm]{ 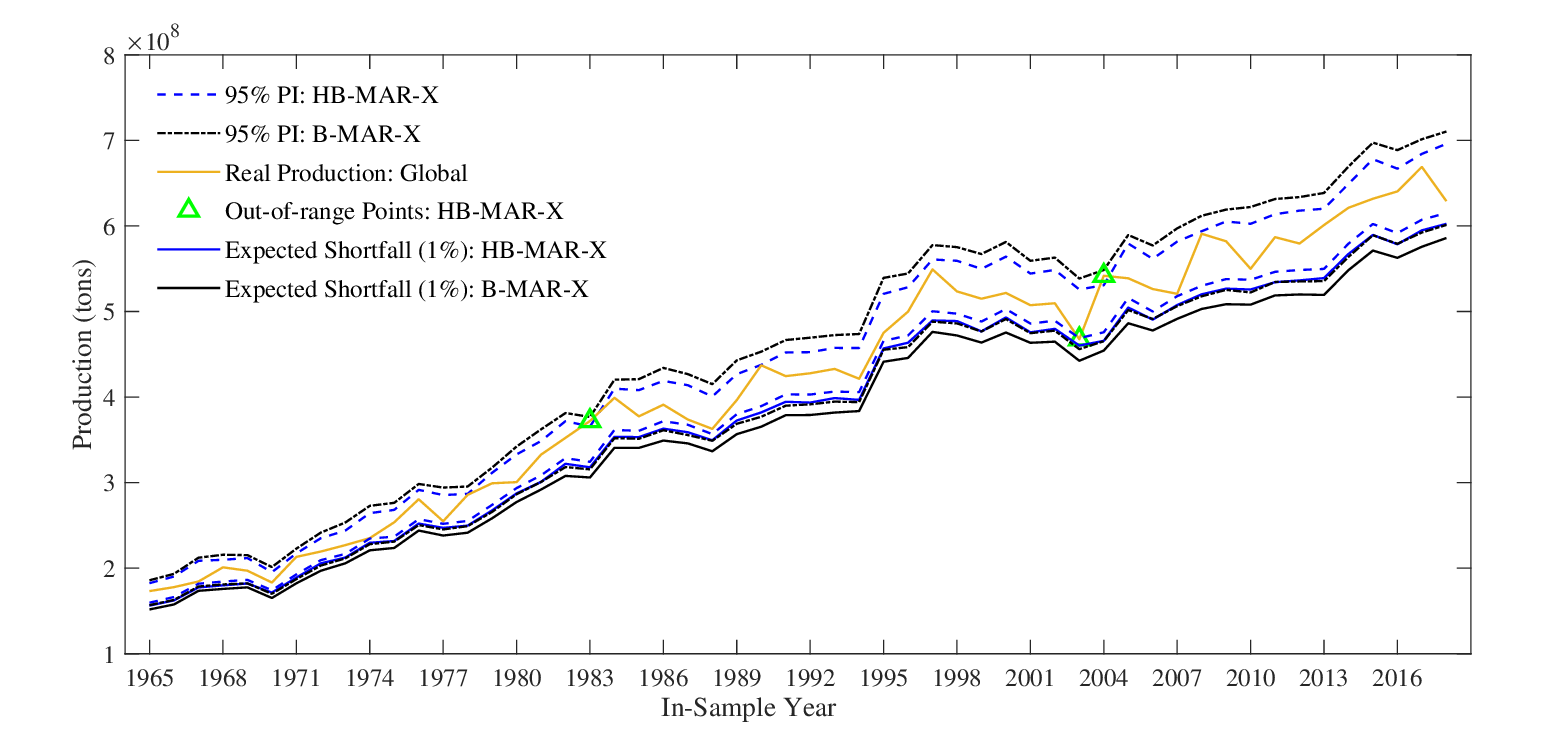}
    \caption{}
    \label{fig:forecast_global_InSample}
  \end{subfigure}


  \begin{subfigure}[b]{1\textwidth}
    \centering
    \includegraphics[width=1\textwidth,height=8cm]{ 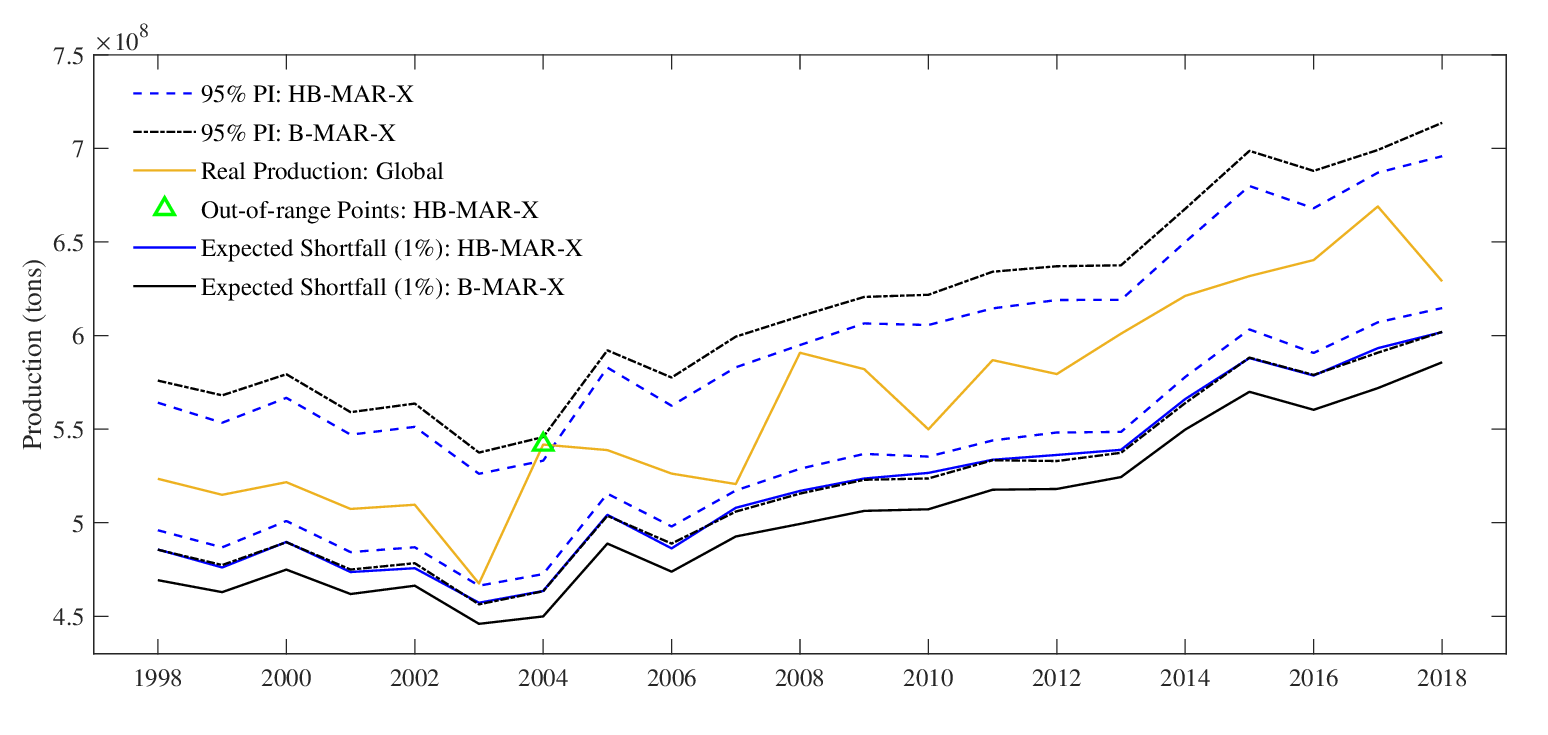}
    \caption{}
    \label{fig:forecast_global_OutSample}
  \end{subfigure}

  \caption{(a) In-sample 95\% prediction interval for global wheat production over all 54 years. 
  (b) Out-of-sample 95\% prediction interval for global wheat production over the 21 evaluation years.}
  \label{fig:forecast_global_combined}
\end{figure}

Figures \ref{fig:forecast_Australia_combined} and \ref{fig:forecast_global_combined} illustrate the in-sample and out-of-sample predictive intervals for wheat yield log returns in Australia and for global wheat production, respectively. 
Across both figures, the predictive intervals produced by the HB-MAR-X model (dashed blue lines) consistently demonstrate better empirical coverage than those of the B-MAR-X model (dashed black lines). During the in-sample period, the empirical coverage rates of the 95\% prediction intervals for Australian wheat yield are approximately 91\% for HB-MAR-X and 72\% for B-MAR-X, while for global wheat production, the corresponding coverage rates are about 94\% for HB-MAR-X and 100\% for B-MAR-X. These results indicate that B-MAR-X tends to underestimate uncertainty for Australian yield returns, while it overestimates uncertainty for global production. This interpretation assumes the superiority of the HB-MAR-X model, which is supported by the higher estimated log evidence reported in Table \ref{tab:log_evidence}.
Notably, the most severe negative log-return years for Australian wheat yields--1967, 1982, 2003, and 2007--coincide with historically documented, severe drought events. These years correspond to most of the observations falling outside the 95\% prediction interval under HB-MAR-X, which is both expected and desirable from a statistical standpoint. A well-calibrated 95\% interval should exclude approximately 5\% of points, and the fact that HB-MAR-X’s misses align closely with known climate-induced yield shocks demonstrates that its predictive uncertainty is appropriately quantified. This alignment confirms that HB-MAR-X captures the underlying yield dynamics effectively.
Furthermore, the HB-MAR-X model enhances tail-risk assessment through its 1\% ES measure (solid blue lines), explicitly quantifying the severity of potential extreme losses. Compared to B-MAR-X, the HB-MAR-X ES consistently suggests deeper potential downside risks. Importantly, the 1\% ES estimates from HB-MAR-X better reflect the severity of potential losses during drought-induced crises such as those in 2003 and 2007. 
This responsiveness allows stakeholders, including agricultural insurers, policymakers, and farmers, to better anticipate and financially prepare for severe yield disruptions due to climatic volatility. Thus, HB-MAR-X’s explicit quantification of extreme downside risk significantly improves agricultural risk management strategies, particularly in enhancing resilience against future climate extremes.

In the out-of-sample evaluation over the 21-year forecasting horizon, the HB-MAR-X model continues to outperform the B-MAR-X model in terms of predictive reliability. For Australian wheat yield log returns, HB-MAR-X achieves an empirical coverage rate of approximately 90\% compared to 71\% for B-MAR-X, again suggesting underestimation of uncertainty by B-MAR-X. Similarly, for global wheat production, HB-MAR-X maintains a more calibrated coverage, while B-MAR-X continues to produce overly wide intervals with full (100\%) empirical coverage, implying an overestimation of predictive uncertainty. 
This contrast is further illustrated by the behavior of the 1\% Expected Shortfall (ES) estimates. In both the in-sample and out-of-sample periods, the ES band from B-MAR-X, represented by the solid black line, lies consistently below that of HB-MAR-X, indicating a more pessimistic view of potential extreme losses. This tendency suggests that B-MAR-X systematically overstates downside risk, which may lead to unnecessarily conservative decision-making and inflated reserve allocations. In contrast, the ES estimates from HB-MAR-X remain more restrained in most years but respond appropriately to genuine global production shocks, such as the sharp decline in 2003.
The HB-MAR-X model’s ability to respond to actual production shocks without overstating risk makes it particularly valuable for managing uncertainty in practical settings. For global stakeholders such as commodity traders, food security agencies, and humanitarian organizations, this translates into clearer signals for when to adjust reserves or implement intervention strategies. Rather than prompting excessive caution in relatively stable years, HB-MAR-X offers a more balanced assessment of risk. This helps decision-makers allocate resources more efficiently while remaining prepared for genuine disruptions driven by climate variability.

\subsection{Probabilistic Yield Projections under Future Climate Scenarios} \label{sec_project}
To comprehensively project future crop yields, the proposed Bayesian hierarchical model integrates climate-driven uncertainties by incorporating projections under representative concentration pathways (RCP2.6, RCP4.5, RCP8.5) and an additional ensemble of 100 carbon concentration pathways generated via QuickClim \citep{kitsios2023machine}. While the uncertainties associated with different climate models and emission scenarios have been previously investigated \citep{li2025machine}, the uncertainty quantification specifically related to the crop modeling framework itself remains unexplored. Our study fills this critical research gap by explicitly quantifying uncertainties arising from the crop model parameters and their interactions with the uncertainty in the climate projections.
The resulting ensemble of predictive distributions allows stakeholders to comprehensively evaluate uncertainties stemming from climate models, emission scenarios, and crop model parameterizations simultaneously. Consequently, this novel approach delivers robust and transparent probabilistic forecasts, substantially advancing beyond traditional deterministic projections (further details are outlined in \ref{appendix:forecasting}).

Moreover, incorporating tail risk measures within our Bayesian hierarchical framework improves risk-informed decision-making by explicitly characterizing extreme yield outcomes, thereby equipping stakeholders with probabilistic projections that reflect a broader range of potential future conditions. This comprehensive uncertainty analysis offers decision-makers a powerful tool for anticipating and managing extreme agricultural risks amid future climate variability and technological change.

\begin{figure}[htbp!]
  \centering
  \captionsetup[subfigure]{skip=2pt}

  \begin{subfigure}[b]{1\textwidth}
    \centering
    \hspace{-1cm}
    \includegraphics[width=1\textwidth,height=8cm]{ 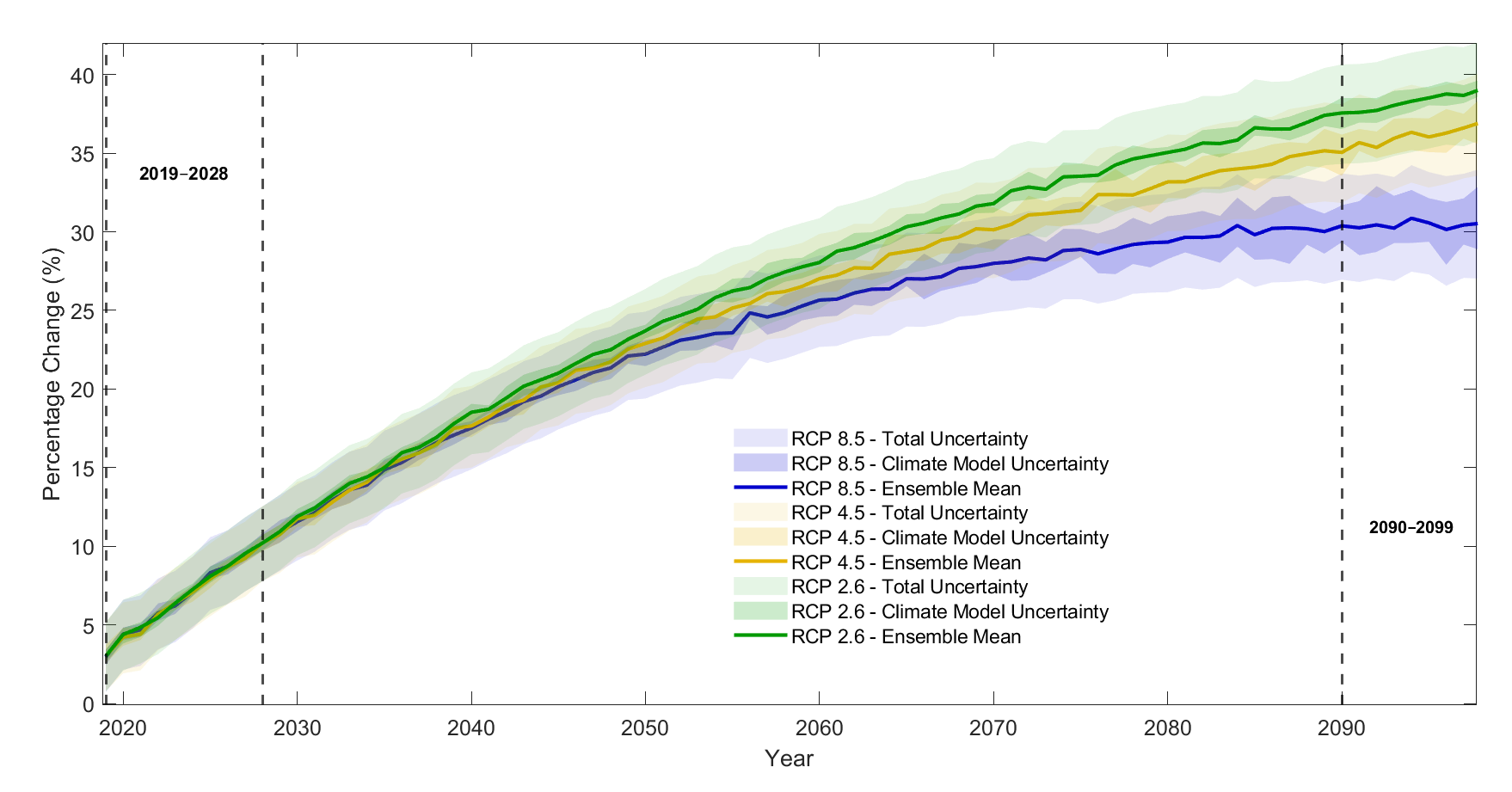}
    \vspace{-1em}
    \caption{}
    \label{fig:Projection_3RCPs}
  \end{subfigure}


  \begin{subfigure}[b]{1\textwidth}
    \centering
    \includegraphics[width=\linewidth]{ 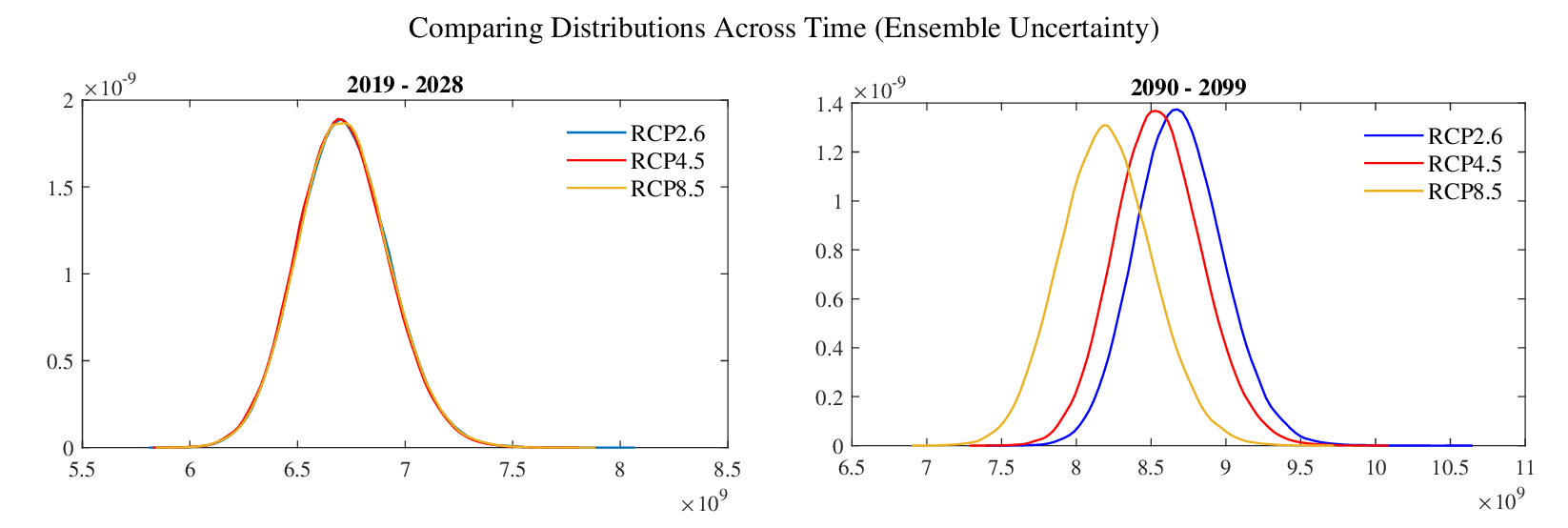}
    \caption{}
    \label{fig:rcp_ensemble_dnt}
  \end{subfigure}


  \begin{subfigure}[b]{1\textwidth}
    \centering
    \includegraphics[width=\linewidth]{ 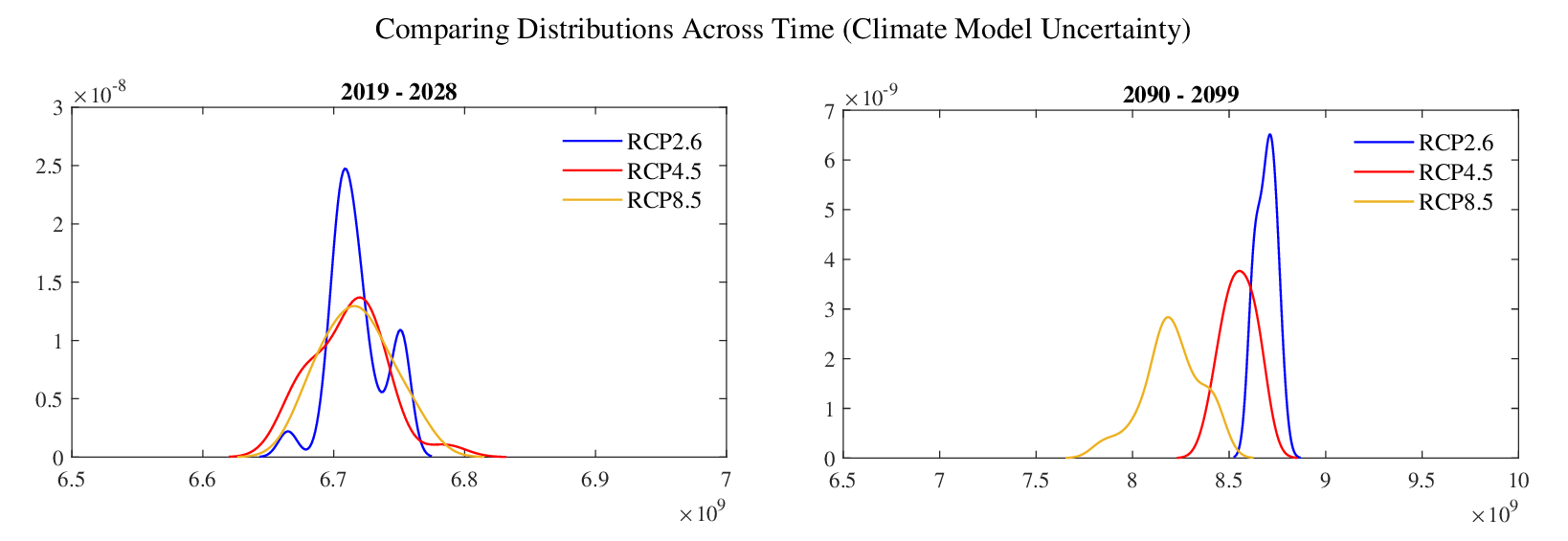}
    \caption{}\vspace{-8pt}
    \label{fig:rcp_model_dnt}
  \end{subfigure}
  
  \caption{(a) Projected percentage change in wheat production, relative to the 2018 reference year, under CMIP5 climate scenarios. 
  (b) Projected probability density of decadal wheat production under future climate scenarios, reflecting ensemble uncertainties from climate models and crop model parameters.
  (c) Projected probability density of decadal wheat production under future climate scenarios, with uncertainties quantified based solely on climate model variability.}
  \label{fig:combined_rcp_projection}
\end{figure}

To quantify the uncertainty attributable to differences among climate models, we first computed the mean projected yield for each model at each time point based on samples from the posterior predictive distribution. We then calculated the 25th and 75th percentiles of these model-specific means to represent the range of yield projections arising from climate model uncertainty. 
To capture the combined uncertainty from both crop model parameters and climate models within each emissions pathway, we aggregated all posterior predictive samples into a single ensemble. For each time point, we pooled the samples from all climate models to form a comprehensive set of ensemble projections. The ensemble mean and interquartile range were then calculated at each time point to summarize the central tendency and spread of projected crop yields, thus reflecting the full distribution of uncertainty across both sources.

Figure \ref{fig:combined_rcp_projection} presents probabilistic projections of decadal wheat production under three representative climate scenarios (RCP2.6, RCP4.5, and RCP8.5), decomposing uncertainty into contributions from climate models, emission scenarios, and crop model parameters. 
Panel (a) displays projected percentage changes in wheat production relative to a 2018 baseline under CMIP5 climate scenarios. All scenarios initially project increasing wheat yields over time, with RCP2.6 showing the highest ensemble mean increase, followed by RCP4.5 and RCP8.5. Notably, the rate of increase gradually flattens over the projection period, especially pronounced under RCP8.5, where projections after around 2080 suggest a possible turning point, potentially indicating a future negative growth rate due to the detrimental impacts of global warming.
Shaded regions quantify different sources of uncertainty, with darker bands representing climate model uncertainty alone, and lighter bands capturing total uncertainty including crop model parameters and climate models. 
The difference between total uncertainty (lighter shaded regions) and climate-model-only uncertainty (darker shaded regions) illustrates the substantial and persistent contribution from crop model parameter uncertainty. This underscores the importance of explicitly quantifying crop model uncertainties alongside climate model variability to inform robust agricultural decision-making.

The predictive probability density of decadal wheat production, accounting for combined climate and crop model uncertainties, is shown in panel (b) of Figure \ref{fig:combined_rcp_projection}. 
In the near term (2019–2028), the distributions across all RCP scenarios are nearly identical, highlighting minimal scenario sensitivity and relatively low uncertainty initially. By the end of the century (2090–2099), substantial differences emerge, with RCP2.6 showing the highest production levels, followed closely by RCP4.5, whereas RCP8.5 indicates considerably lower yields. Moreover, RCP8.5 exhibits the widest distribution, reflecting increased uncertainty and greater exposure to adverse agricultural outcomes under extreme warming scenarios. 
Panel (c) isolates the uncertainty arising solely from climate model variability, holding crop model parameters constant within each emission scenario. In contrast to panel (b), which integrates uncertainty from both climate models and crop model parameters, panel (c) exhibits noticeably narrower distributions.
This comparison clearly highlights that a substantial portion of total uncertainty in wheat production projections stems from crop model parameters. 
Collectively, the contrast between panels (b) and (c) underscores the critical value of our Bayesian hierarchical modeling framework. By explicitly decomposing multiple uncertainty sources, this approach significantly improves the transparency and robustness of crop yield projections, thus enhancing long-term agricultural planning and risk management under climate change and technological uncertainty.

\begin{figure}[htbp!]
  \centering
  \begin{subfigure}[b]{0.8\textwidth}
    \centering
    \hspace{-2.75cm}
    \includegraphics[width=1.1\textwidth,height=8.5cm]{ 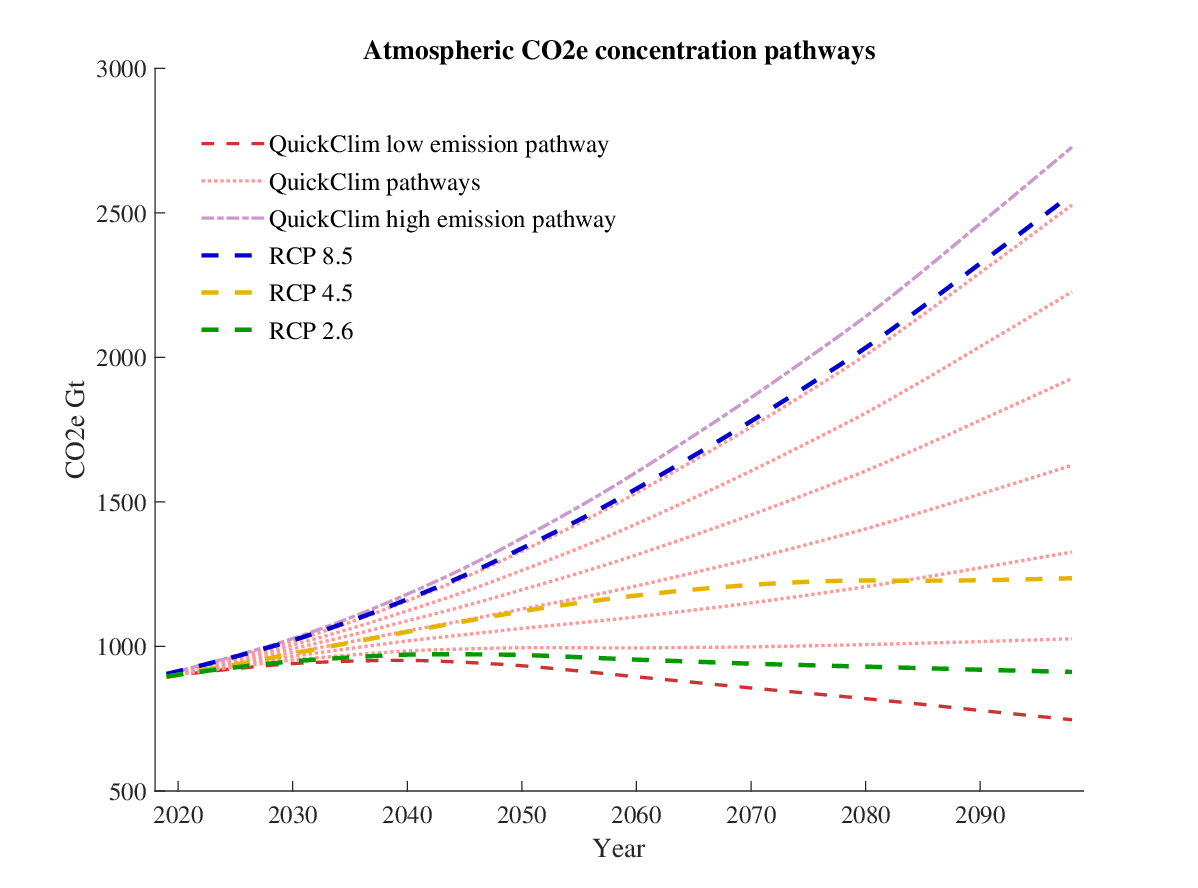}
    \label{fig:Projection_QC_emission}
    \caption{}
  \end{subfigure}

  \vspace{1em}

  \begin{subfigure}[b]{0.8\textwidth}
    \centering
    \hspace{-2.75cm}
    \includegraphics[width=1\textwidth,height=7.75cm]{ 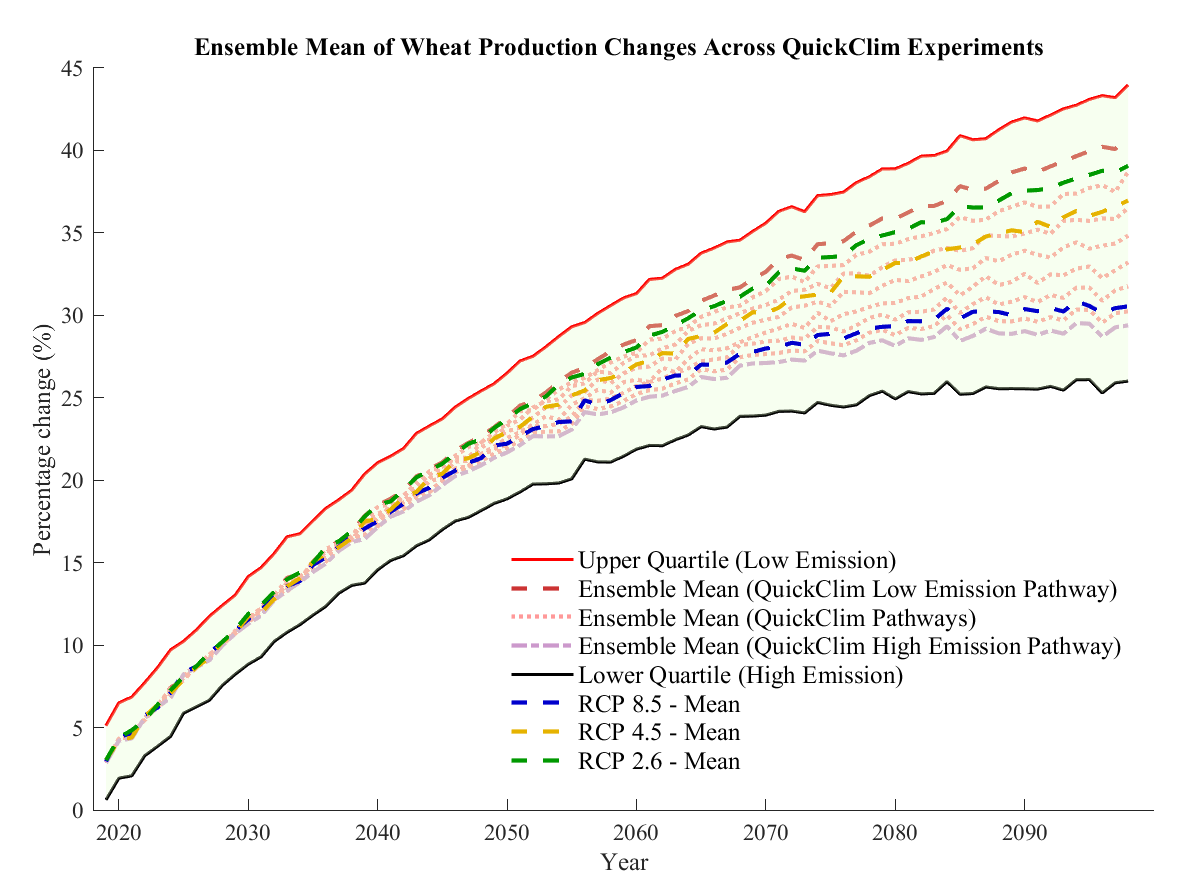}
    \label{fig:Projection_QC}
    \caption{}
  \end{subfigure}
  
  \caption{(a) Projected atmospheric carbon dioxide equivalent ($CO_{2e}$) emission pathways derived from the QuickClim approach (showing the lowest, the highest, and every 15th of 100 ensemble members) and from the standard representative concentration pathways (RCP2.6, RCP4.5, and RCP8.5). (b) Projected percentage change in wheat production relative to the 2018, under the same subset of QuickClim $CO_{2e}$ pathways and RCPs. Quartile bounds for the QuickClim ensemble are also shown.}
  \label{fig:Combined_Figure_QC}
\end{figure}

Panel (a) of Figure \ref{fig:Combined_Figure_QC} displays projected atmospheric carbon dioxide equivalent ($CO_{2e}$) concentrations, associated with the QuickClim approach and the standard RCPs. For clarity, only a subset of the 100 QuickClim emission pathways is shown: specifically, the lowest, the highest, and every 15th pathway in the ensemble. These are depicted alongside RCP 2.6, RCP 4.5, and RCP 8.5 (shown as dashed green, yellow, and blue lines, respectively). The displayed pathways still encompass the full range of outcomes, from the most optimistic (lowest emission) to the most pessimistic (highest emission), with representative intermediate scenarios providing additional context. Notably, the QuickClim high emission scenario exceeds RCP 8.5 by end of the century, while the low emission scenario falls below RCP 2.6, indicating a broader range than the standard RCPs. 
Panel (b) presents the projected percentage changes in wheat production relative to the 2018 baseline, based on the 100 QuickClim $CO_{2e}$ emission pathways, a subset of which is shown in panel (a). The corresponding ensemble mean trajectories are shown as dotted lines. For comparison, ensemble means under standard RCP scenarios (RCP 2.6, 4.5, and 8.5) are shown as dashed green, yellow, and blue lines, respectively.
Two highlighted dashed lines represent wheat production outcomes under the lowest (red dashed) and highest (purple dashed) QuickClim emission pathways, offering direct insights into the upper and lower bounds of projected yield responses. Additionally, the shaded envelope defined by the upper quartile (solid red) and lower quartile (solid black) captures the ensemble uncertainty across all QuickClim experiments.
Overall, the rate of increase in wheat production varies substantially depending on the $CO_{2e}$ pathway. While lower-emission scenarios show steady and sustained gains in yield, some higher-emission scenarios exhibit a decreasing growth rate over time, suggesting diminishing benefits or potential negative effects under more extreme climate conditions. 
Importantly, the inclusion of ensemble spread and quartile bounds in panel (b) emphasizes the degree of uncertainty in future wheat production projections. 
The results highlight that, after factoring in assumed technological development, moderate increases in $CO_{2e}$ and low levels of warming may enhance wheat yields in certain regions via fertilization effects. However, as demonstrated in the work of \citet{li2025machine}, these climate-related benefits are largely confined to a small number of countries under low-emission scenarios. Most projected global yield increases are attributable to technological progress rather than climate change itself. 
These insights underscore the importance of distinguishing between technological and climate-driven contributions to yield outcomes, and highlight the need for robust adaptation planning across a range of plausible climate futures. 
They also highlight the importance of incorporating a wide range of plausible climate and development trajectories into impact assessments and decision-making frameworks for agricultural adaptation.

\begin{figure}[htbp!]
  \centering
  \hspace{-2cm}
\includegraphics[width=1\textwidth,height=9.25cm]{ 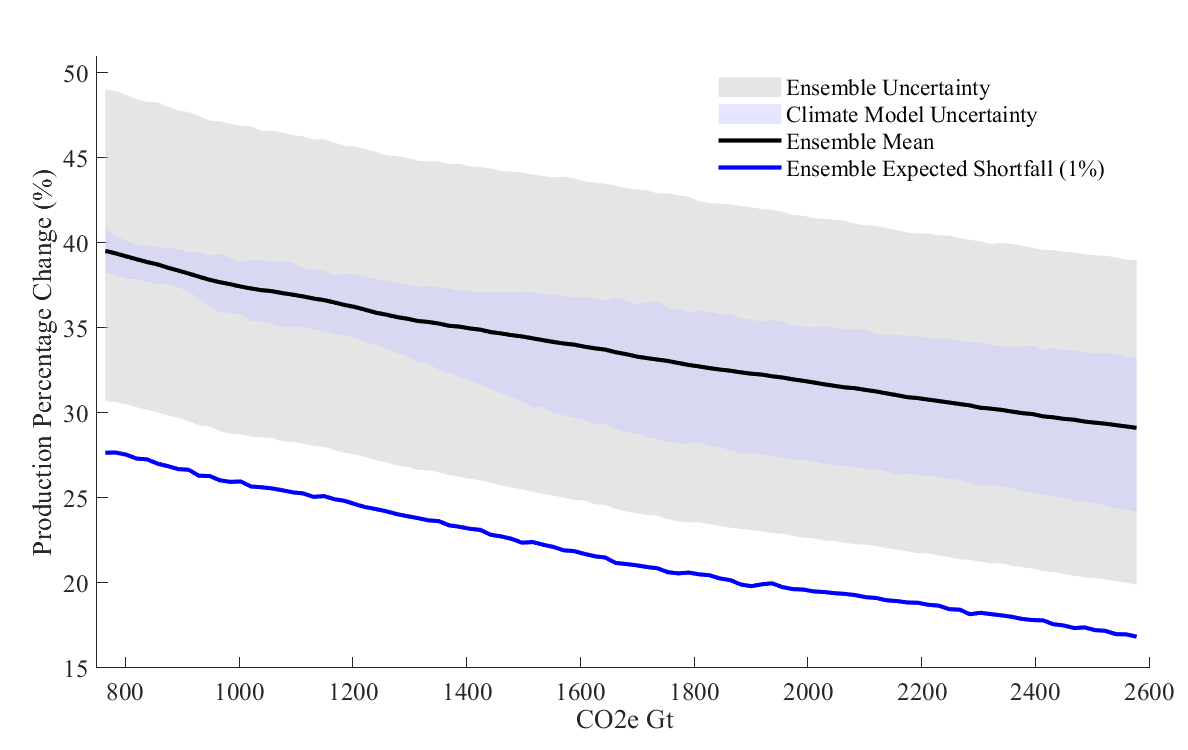} 
  \caption{Projected percentage change in wheat production under 100 atmospheric carbon dioxide equivalent ($CO_{2e}$) concentration pathways, based on the QuickClim approach. Projections are relative to the 2018 baseline and represent the average of the final decade. The black line denotes the mean values derived from 18 different climate models. The inner shaded area represents the 95\% prediction interval, capturing uncertainty from climate models, while the outer shaded area accounts for uncertainties from both climate and agricultural models. The bottom blue edge line represents the 1\% ES of the change in wheat production.}

  \label{fig:Projection_QC_decade}
\end{figure}

Figure \ref{fig:Projection_QC_decade} displays projected percentage changes in wheat production as a function of atmospheric $CO_{2e}$ concentrations, based on 100 emission pathways generated by the QuickClim approach. All projections are relative to a 2018 baseline and represent averages for the final decade of the 21st century. 
The solid black curve shows the ensemble-mean response obtained from simulations with 18 climate models coupled to the proposed crop model. The mean declines almost linearly, demonstrating that progressively higher long-run emissions are associated with progressively smaller yield gains. Uncertainty is represented by two nested shaded bands. The inner blue band reflects the 2.5th–97.5th percentile range associated with climate model spread alone, while the outer grey band incorporates additional uncertainty from crop model responses. Both intervals widen gradually as $CO_{2e}$ increases. 
This trend suggests that as emissions rise, not only do mean projections decline, but the spread of possible outcomes also broadens. 
The blue curve at the bottom represents the 1\% expected shortfall, defined as the mean of the most adverse 1\% of projected outcomes. This metric declines from approximately 28\% to 17\% across the range of $CO_{2e}$ concentrations. The deepening lower tail indicates that while central estimates of production remain positive, the worst-case outcomes become more severe under higher-emission futures. 
By way of comparison, the global population was approximately 7.6 billion in 2018 \citep{PRB2018}. Under the United Nations medium-variant scenario, it is expected to rise to about 10.4 billion in the mid-2080s and then plateau at roughly that level through 2100 \citep{UN2022}. This corresponds to an approximate 36\% increase from 2018 to 2100. Although these population projections refer to individual annual estimates, rather than decade averages as used for wheat production projections (2090–2099), the comparison illustrates the potential magnitude of future challenges in meeting global wheat demand under higher-emission scenarios.
Together, these results indicate that increasing cumulative $CO_{2e}$ concentrations lead to reduced expected gains in wheat production, greater overall uncertainty, and heightened risk of extreme negative outcomes. These findings underscore the importance of incorporating both central tendencies and tail risks when evaluating the agricultural consequences of long-term emission pathways.

\section{Discussion and Conclusion}
\label{sec_discussion}



This study presents a new Bayesian hierarchical framework, called HB-MAR-X, designed to improve the way we forecast global crop yields, with a focus on wheat. By explicitly modelling country-specific variances and integrating uncertainties from climate models, emission scenarios, and crop model parameters, our approach addresses key limitations of traditional pooled-error models. The HB-MAR-X framework demonstrates improved calibration and probabilistic accuracy for both in-sample and out-of-sample predictions, particularly during periods marked by extreme yield shocks, such as major drought events. This enhances the reliability of tail risk metrics, such as Value-at-Risk (VaR) and Expected Shortfall (ES), thereby supporting more effective and timely risk management and policy responses.

A key strength of the proposed framework is its transparent decomposition of predictive uncertainty into contributions from climate models, emissions pathways, and crop model parameters. Our results show that, while climate model uncertainty remains the dominant source of forecast spread, crop model parameter uncertainty makes a substantial and persistent contribution to the total uncertainty throughout the projection period. This is evident in the marked widening of predictive intervals when crop model parameter uncertainty is included, particularly toward the end of the century. Consequently, comprehensive risk assessment and robust decision-making require that both sources of uncertainty are explicitly quantified and communicated. 

Furthermore, projections across a wide range of emission pathways reveal that higher emissions are associated with lower ensemble mean yields, broader predictive intervals, and a deepening of downside risk, as measured by the 1\% ES. This highlights the increasing vulnerability of global food production to adverse extremes under high-emission scenarios and reinforces the need for adaptation and mitigation strategies that explicitly consider tail risks, not just central estimates.


In addition to tail-specific metrics, our Bayesian framework supports a full evaluation of expected climate risk under each emissions scenario $s$, following recent advances in the literature \citep{villalta2020spatio}. The unconditional expected loss can be expressed as $\mathbb{E}[\text{Loss} | s] = \int_h \int_v e \, v \, p(v | h) \, p(h | s) \, dv \, dh = \mathbb{E}[e] \int E[v | h] \, p(h | s) \, dh$, where $p(h | s)$ represents the probability density of a climate hazard (e.g., temperature or precipitation anomalies) under scenario $s$, informed by emulators such as QuickClim or CMIP ensembles; $p(v | h)$ denotes the vulnerability distribution--the conditional yield loss given a hazard $h$; and $e$ represents the value of the exposed assets (e.g., production area or value), factored out as its mean $\mathbb{E}[e]$.
Our Bayesian posterior predictive samples naturally produce draws from $p(v | h)$ while fully propagating uncertainty in the crop model parameters. As a result, we can estimate not only the mean expected loss but also uncertainty intervals for $\mathbb{E}[\text{Loss} | s]$, alongside previously reported tail risk measures such as ES. This unified probabilistic framework provides decision-makers with a richer and more robust picture of future climate risks, integrating both central and extreme outcomes. 
By contrast, the OLS MAR-X model \citep{li2025machine} yields a deterministic vulnerability function $\hat{v}(h)$, which when applied to the expected loss framework results in $\mathbb{E}[\text{Loss} | s]_{\text{OLS}} = \mathbb{E}[e] \int \hat{v}(h) \, p(h | s) \, dh.$ This approach does not account for parameter uncertainty in $\hat{v}$ beyond its expected value, and thus likely underestimates both the variability in losses and the severity of tail outcomes. 
In contrast, our HB-MAR-X model treats crop-model parameters $\theta$ as random, and the full expected-loss integral under our framework
captures the complete uncertainty structure of the vulnerability–hazard relationship. This provides not only more realistic central estimates, but also probabilistic bounds on expected losses, which are critical for robust climate risk management and adaptation planning.


In summary, the HB-MAR-X model offers a significant methodological advance in crop yield forecasting by delivering robust, interpretable, and probabilistically coherent projections that integrate multiple sources of uncertainty. These results equip policymakers and stakeholders with the risk-sensitive information necessary to enhance resilience and adaptability in agricultural planning under climate change. Looking ahead, future research should focus on extending hierarchical Bayesian methodologies to additional crops and process-based models, as well as integrating yield forecasts with economic models to more comprehensively assess the broader impacts of biophysical risks on global food systems.


\section*{Reference}
\renewcommand{\baselinestretch}{1}
\setlength{\parskip}{-1em}
\renewcommand\refname{}
\makeatletter
\renewcommand\@biblabel[1]{}
\makeatother
\bibliographystyle{apalike}
\bibliography{ref_quickAgri}

\renewcommand{\baselinestretch}{1.5}
\setlength{\abovedisplayskip}{4.5pt}
\setlength{\belowdisplayskip}{4.5pt}
\setlength{\parskip}{0.8em}

\renewcommand{\thefigure}{A.\arabic{figure}}
\renewcommand{\thetable}{A.\arabic{table}}
\setcounter{figure}{0}
\setcounter{table}{0}

\appendix
\setcounter{section}{0}
\renewcommand{\thesection}{Appendix~\arabic{section}}
\section{Sequential Monte Carlo} \label{SMC}

We use Sequential Monte Carlo (SMC) methods for static Bayesian models \citep{chopin2002sequential} to estimate our proposed hierarchical crop yield framework. SMC can be viewed as an extension of Importance Sampling (IS) \citep{neal2001annealed}, adapted to a sequential setup that allows for more effective sampling from complex posteriors. 

In a standard IS approach, samples $\{\bm{\theta}^i\}_{i=1}^N$ are drawn from a convenient proposal $g(\bm{\theta})$, and each sample is assigned a weight 
\[
w^i = \frac{f(\bm{y}\mid\bm{\theta}^i)\,\pi(\bm{\theta}^i)}{g(\bm{\theta}^i)}, \quad
W^i = \frac{w^i}{\sum_{j=1}^N w^j},
\]
so that the weighted sample, $\{W^i, \bm{\theta}^i\}_{i=1}^N$, approximates the posterior $\pi(\bm{\theta}\mid\bm{y})$. The efficiency of these weights can be gauged via the effective sample size (ESS) \citep{liu2008monte}, defined as:
\[
\text{ESS} = \frac{1}{\sum_{i=1}^N (W^i)^2}.
\]
The ESS ranges from 1 to $N$, and reaches its maximum when all weights are equal, indicating an ideal match between the importance distribution and the target distribution. 

SMC generalizes IS by introducing a sequence of intermediate distributions,
\[
\pi_t(\bm{\theta}\mid\bm{y}) \quad \text{for } t=0,\dots,T,
\]
that transitions smoothly from a simpler reference (often the prior $\pi(\bm{\theta})$) to the full posterior. A commonly used approach is likelihood annealing, where the likelihood function is gradually introduced through a sequence of tempering parameters $\gamma_t$, evolving from $\gamma_0=0$ to $\gamma_T=1$. Specifically, at iteration $t$, the intermediate distribution takes the form 
\[
    \pi_t(\bm{\theta}\mid\bm{y}) \;\propto\; f(\bm{y}\mid\bm{\theta})^{\,\gamma_t}\,\pi(\bm{\theta}),
\]
thus allowing the posterior distribution to be approximated progressively.
At each iteration $t$, we reweight particles according to the ratio of target densities, possibly resample to discard low‑weight particles, and then apply an MCMC move step to rejuvenate diversity in the particle set. This ensures a robust representation of each intermediate distribution.


After resampling, an MCMC move step is applied to each particle to mitigate degeneracy and enhance exploration of the target distribution \citep{chopin2002sequential}. In this step, an MCMC kernel proposes a new particle, $\bm{\theta}_t^*$, for each current particle, $\bm{\theta}_t^i$, using a proposal distribution $q(\bm{\theta}_t^*|\bm{\theta}_t^i)$. We employ a multivariate normal random walk kernel:
\[
q(\bm{\theta}_t^*|\bm{\theta}_t^i) = \mathcal{N}\left(\bm{\theta}_t^i,\, h^2\,\hat{\Sigma}_t\right),
\]
where $\hat{\Sigma}_t$ is the empirical covariance matrix of the particle set at iteration $t$, and $h$ is a scaling factor determined adaptively. The proposed move is accepted with probability
\[
\alpha\left(\bm{\theta}_t^i \rightarrow \bm{\theta}_t^*\right) = \min\left(1, \, \frac{\pi_t(\bm{\theta}_t^*|\bm{y})\,q(\bm{\theta}_t^i|\bm{\theta}_t^*)}{\pi_t(\bm{\theta}_t^i|\bm{y})\,q(\bm{\theta}_t^*|\bm{\theta}_t^i)} \right),
\]
ensuring that the updated particle set remains consistent with the intermediate target distribution $\pi_t(\bm{\theta}|\bm{y})$. To optimize particle movement, we tune the scale $h$ based on the expected square jumping distance (ESJD; \citealp{sherlock2009optimal, salomone2018unbiased}), defined as
\[
\text{ESJD} = \alpha\left(\bm{\theta}_t^i \rightarrow \bm{\theta}_t^*\right) \cdot \|\bm{\theta}_t^i - \bm{\theta}_t^*\|_{\hat{\Sigma}_t}^2,
\]
where $\|\bm{\theta}_t^i - \bm{\theta}_t^*\|_{\hat{\Sigma}_t}$ denotes the Mahalanobis distance between the current and proposed particles. The optimal scale is selected by comparing the median ESJD across particles for a range of candidate scales, and choosing the scale that maximizes this median. Once the optimal scale is determined, the MCMC kernel is applied for adaptively chosen number of iterations, $R_t$, until the ESJD of a specified proportion of particles exceeds a predetermined threshold. This adaptive move step ensures robust local exploration while maintaining the integrity of the intermediate distribution.


In our hierarchical crop yield setting, SMC offers several key benefits. It provides robust exploration by navigating the potentially high‑dimensional posterior space of crop yield parameters (including country‑level variances) more effectively than a single‑stage sampler that attempts to sample all parameters simultaneously. Additionally, by gradually updating from a simpler distribution to the posterior, the algorithm achieves adaptive accuracy, reducing the risk of particle impoverishment and local trapping. Finally, SMC naturally facilitates Bayesian model selection and predictive distributions through its evolving particle set, thereby offering a unified framework for posterior inference and out‑of‑sample yield prediction.

Algorithm~\ref{SMC1} describes the key steps of our SMC procedure in detail. This method underpins the Bayesian estimation of our proposed hierarchical crop yield model, and it also facilitates robust model selection and prediction, as detailed in subsequent sections. SMC’s efficiency and advantages in various econometric contexts have been noted elsewhere \citep{li2021efficient, li2023bayesian}, making it a suitable choice for our application.

\begin{algorithm}[H]  
\SetAlgoLined
\caption{SMC Sampler}
\KwIn{Observations $\bm{y}$, Number of particles $N$, Threshold $\alpha$}
\KwOut{Particle set $\left\{W_t^i, \bm{\theta}_t^i\right\}_{i=1}^N$}
\For{$i=1$ \KwTo $N$}{
    Sample $\bm{\theta}_0^i \sim \pi(\cdot)$\;
    Set $W_0^i = 1/N$\;
}
\For{$t=1$ \KwTo $T$}{
    Reweight: $w_t^i = W_{t-1}^i \frac{\eta_t(\bm{\theta}_{t-1}^i|\bm{y})}{\eta_{t-1}(\bm{\theta}_{t-1}^i|\bm{y})}$\;
    Set $\bm{\theta}_t^i = \bm{\theta}_{t-1}^i$ for $i=1, \dots, N$\;
    Normalize weights: $W_t^i = w_t^i / \sum_{k=1}^N w_t^k$ for $i=1, \dots, N$\;
    Compute ESS: $\text{ESS} = 1 / \sum_{i=1}^N (W_t^i)^2$\;
    
    \If{$\text{ESS} < \alpha N$, with $\alpha \in (0,1]$}{
        Resample the particles and set $W_t^i = 1/N$\; 
        Compute $\hat{\Sigma}_t$ from the particle set $\left\{\bm{\theta}_t^i\right\}_{i=1}^N$\;
        Compute the median Mahalanobis distance $D_\mathrm{desired}$\;
        \While{Number of particles with ESJD $>$ $D_\mathrm{desired}$ is below target}{
            Move each particle using MCMC kernel with proposal $q(\bm{\theta}_t^{i,*}|\bm{\theta}_t^i)$\;
        }
    }
}
\label{SMC1}
\end{algorithm}

\section{Bayesian Model Selection} \label{sec:model_selection}
In practice, it is often important not only to analyze each model individually but also to compare the performance of different models against one another. In a Bayesian framework, this comparison is facilitated by evaluating the posterior probabilities of the models. Suppose there are \( M \) candidate models, labeled as \( m = 1, \ldots, M \). The posterior probability of model \( m \), denoted as \( P(m \mid \bm{y}) \), can be computed using Bayes' theorem:
\[
P(m \mid \bm{y}) = \frac{P(\bm{y} \mid m) P(m)}{P(\bm{y})}.
\]
Here, \( P(\bm{y} \mid m) \) represents the marginal likelihood (or model evidence), which is the integral of the likelihood over the parameter space of the model. The marginal likelihood is expressed as:
\[
P(\bm{y} \mid m) = \int P(\bm{y} \mid \bm{\theta}_m, m) P(\bm{\theta}_m \mid m) \, \mathrm{d} \bm{\theta}_m.
\]
To compare two models, \( m_1 \) and \( m_2 \), the posterior odds ratio can be used:
\[
\frac{P(m = m_1 \mid \bm{y})}{P(m = m_2 \mid \bm{y})} = \frac{P(\bm{y} \mid m = m_1)}{P(\bm{y} \mid m = m_2)} \cdot \frac{P(m = m_1)}{P(m = m_2)}.
\]
If equal prior probabilities are assumed, i.e., \( P(m = m_1) = P(m = m_2) \), the posterior odds ratio is equivalent to the ratio of the marginal likelihoods, which is called the Bayes factor (\citealp{jeffreys1935some,kass1995bayes}):
\[
\frac{P(m = m_1 \mid \bm{y})}{P(m = m_2 \mid \bm{y})} = \frac{P(\bm{y} \mid m = m_1)}{P(\bm{y} \mid m = m_2)}.
\]

While the marginal likelihood provides a basis for model selection, its calculation often involves challenging integrals that can be computationally intensive. A summary of common approaches to estimate marginal likelihoods is available in \citet{friel2012estimating}.
SMC methods offer an efficient means of estimating model evidence as a by-product of the sampling process. The total evidence \( Z \) is estimated incrementally as:
\[
Z = \frac{Z_T}{Z_0} = \prod_{t=1}^T \frac{Z_t}{Z_{t-1}},
\]
where \( Z_0 = 1 \) \citep{del2006sequential}. In the context of likelihood annealing SMC, the ratio of normalizing constants at iteration \( t \), \( Z_t / Z_{t-1} \), is given by:
\[
\frac{Z_t}{Z_{t-1}} = \int_{\bm{\theta}} f(\bm{y} \mid \bm{\theta})^{\gamma_t - \gamma_{t-1}} \pi_{t-1}(\bm{\theta} \mid \bm{y}) \, \mathrm{d} \bm{\theta}.
\]
The log-evidence, \( \log Z \), can be approximated by:
\[
\log Z \approx \sum_{t=1}^T \log \left( \sum_{i=1}^N w_t^i \right),
\]
where \( w_t^i \) denotes the unnormalized weights from Algorithm \ref{SMC1}. 
Thus, the computed evidence \( Z \) serves as the marginal likelihood \( P(\bm{y} \mid m) \) in the model comparison framework, providing the necessary basis for calculating Bayes factors between competing models.

\section{Bayesian Forecasting Details}
\label{appendix:forecasting}
This appendix provides supplementary methodological details referenced in the main text regarding the Bayesian forecasting procedure.

\subsection*{Bayesian Framework for Posterior Predictive Distributions}
Within a fully Bayesian framework, it is possible to generate the posterior predictive distribution for upcoming observations. Given the existing dataset $\bm{y}_{1:t}$ up to time $t$, the posterior predictive distribution for new values $\hat{\bm{y}}_{t+1}$ is formally given by:
\begin{equation}
\mathrm{P}(\hat{\bm{y}}_{t+1}|\bm{y}_{1:t}) = \int_{\bm{\theta}} \mathrm{P}(\hat{\bm{y}}_{t+1}|\bm{\theta}, \bm{y}_{1:t}) \pi_t(\bm{\theta}|\bm{y}_{1:t}) \, \mathrm{d}\bm{\theta}.
\end{equation}
With the SMC method, this posterior predictive distribution becomes straightforward to estimate. Let the weighted sample from the posterior $\pi_t(\bm{\theta}|y_{1:t})$ be denoted by $\left\{W_t^i, \bm{\theta}_t^i\right\}_{i=1}^N$. Then, an approximation of the posterior predictive distribution can be represented by $\left\{W_t^i,\hat{\bm{y}}_{t+1}^i\right\}_{i=1}^N$, where each $\hat{\bm{y}}_{t+1}^i$ is sampled from the distribution $p(\bm{y}_{t+1}|\bm{y}_{1:t}, \bm{\theta}_t^i)$ for \(i = 1, \dots, N\).
Using this weighted set, point estimates, such as the mean or median, and predictive intervals are straightforward to compute. 

In the case of our hierarchical crop production model, the key parameters for prediction are those that directly affect the likelihood, namely the model parameters $\bm{\Theta} = \{\alpha, \lambda, \theta_1, \theta_2, \theta_3, \theta_4, \sigma_i^2 \mid i = 1,\dots, K\}$.
The posterior predictive distribution for a future observation \(\hat{\bm{y}}_{t+1}\) is obtained by integrating over these parameters:
\begin{align}
    \mathrm{P}(\hat{\bm{y}}_{t+1} \mid \bm{y}_{1:t}) = \int \mathrm{P}(\hat{\bm{y}}_{t+1} \mid \bm{\Theta}, \bm{y}_{1:t})\, \pi_t(\bm{\Theta} \mid \bm{y}_{1:t}) \, \mathrm{d}\bm{\Theta}.
\end{align}
Using the SMC method, the posterior \(\pi_t(\bm{\Theta} \mid \bm{y}_{1:t})\) is approximated by the weighted sample \(\{W_t^i, \bm{\Theta}_t^i\}_{i=1}^N\). 
Consequently, the predictive distribution is approximated by the weighted set \(\{W_t^i, \hat{\bm{y}}_{t+1}^i\}_{i=1}^N\).

\subsection*{Predictive Accuracy: Leave-Future-Out Cross-Validation (LFO-CV)}
To evaluate the predictive accuracy of distributions obtained from various models, we utilize the leave-future-out cross-validation (LFO-CV) method, as proposed by \citet{burkner2020approximate}. Traditional leave-one-out cross-validation (LOO-CV) inadequately addresses sequential data, potentially introducing predictive bias due to ignoring temporal structure. By contrast, LFO-CV appropriately handles the sequential nature of the data, where the predictive accuracy is quantified using the expected log predictive density (elpd; \citealp{vehtari2017practical}), which is approximated as follows:
\begin{align}
\mathrm{elpd}= \sum_{t=\tau-1}^{T-1}\log{p(\bm{y}_{t+1}|\bm{y}_{1:t})},
\end{align}
where $\tau$ represents the point in time from which we begin evaluating predictions, and $T$ is the total number of observations in the dataset \citep{burkner2020approximate}. 
The density $p(\bm{y}_{t+1}|\bm{y}_{1:t})$ is estimated via kernel density estimation based on the weighted sample $\left\{W_t^i,\hat{\bm{y}}_{t+1}^i\right\}_{i=1}^N$, evaluated at the true value $\bm{y}_{t+1}$. 
Our focus on the elpd stems from its ability to evaluate the entire predictive distribution rather than a single point estimate (such as the mean or median). This comprehensive evaluation is particularly advantageous from a Bayesian perspective, as it captures the uncertainty in the predictive performance more effectively.

\subsection*{Ensemble Probabilistic Projections Under Climate Uncertainty}

To forecast future crop yields, the trained model is applied to projected climate variables under various climate change scenarios. 
For each climate scenario and climate model $\mathrm{C_M}$, the posterior predictive distribution for future observations \(\hat{\bm{y}}_{t+1}\) is given by integrating across model parameters $\bm{\Theta}$:
\begin{equation}
    \mathrm{P}(\hat{\bm{y}}_{t+1} \mid \bm{y}_{1:t}, \mathrm{C_M}) = \int \mathrm{P}(\hat{\bm{y}}_{t+1} \mid \bm{\Theta}, \bm{y}_{1:t}, \mathrm{C_M})\, \pi_t(\bm{\Theta} \mid \bm{y}_{1:t}, \mathrm{C_M}) \, \mathrm{d}\bm{\Theta}.
\end{equation}
The collection \(\{W_t^i, \hat{\bm{y}}_{t+1}^i\}_{i=1}^N\) serves as an empirical approximation of the full predictive distribution. 

The resulting ensemble of crop yield projections, which incorporates forecasts from the RCPs as well as additional QuickClim-generated pathways, provides comprehensive information for decision-making. Notably, our approach integrates uncertainties from multiple sources—encompassing differences among climate models, variations across climate scenarios, and uncertainties in crop model parameters—to deliver a complete probabilistic assessment of future crop yields. This innovative ensemble framework enables stakeholders to evaluate a wide range of future scenarios, with full transparency regarding the inherent uncertainties. This integrated uncertainty analysis represents a substantial advancement over traditional methods, providing decision-makers with a robust tool for navigating the complex uncertainties of future agricultural production.

\FloatBarrier
\section*{Extended Data Figures Tables}\label{appendix_A}

\begin{figure}[htbp]
  \centering
  \vspace{-0.2cm}
\includegraphics[width=0.9\textwidth,height=10cm]{ 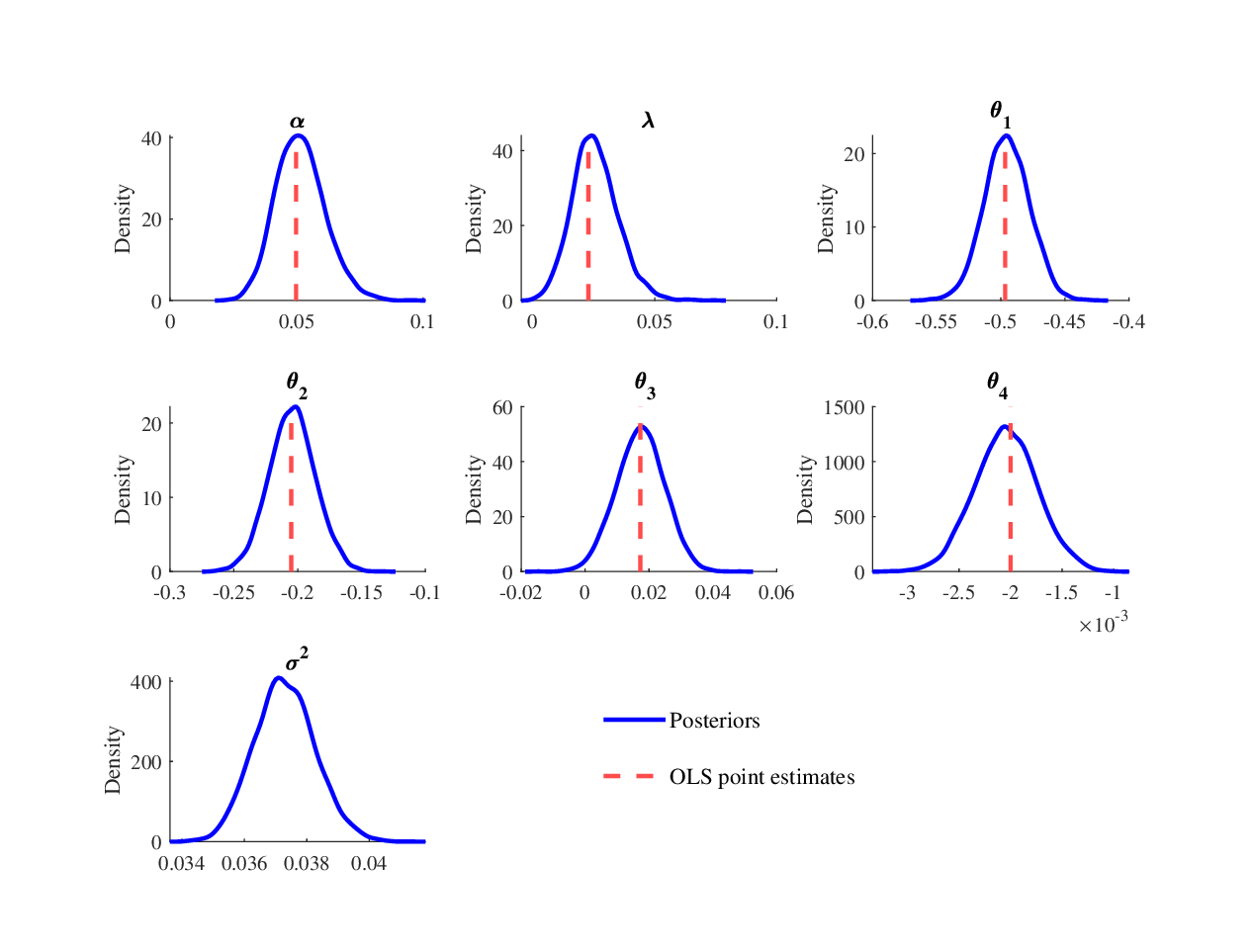} 
\vspace{-1cm}
  \caption{Marginal posterior distributions of parameters from the Bayesian MAR-X model with exponentially decaying intercepts, with point non-linear least squares estimates of Model \ref{model_TVI_2} indicated by dashed vertical lines.}

  \label{fig:posterior_marx}
\end{figure}

\begin{table}[h!]
\centering
\caption{List of Wheat Growing Countries} 
\label{crop_countries}
\begin{tabular}{p{10cm} p{10cm}}
\hspace*{-1.75cm}
    \centering
    \begin{tabular}{|p{4cm}|p{4cm}|p{4cm}|}
    \hline
    \multicolumn{3}{|c|}{{Africa}} \\
    \cellcolor{gray!20}\textbf{Algeria} & Angola & Egypt \\
    \cellcolor{gray!20}\textbf{Kenya} & \cellcolor{gray!20}\textbf{Lesotho} & \cellcolor{gray!20}\textbf{Libya} \\
    \cellcolor{gray!20}\textbf{Morocco} & \cellcolor{gray!20}\textbf{Mozambique} & Namibia \\
    \cellcolor{gray!20}\textbf{South Africa} & \cellcolor{gray!20}\textbf{Tanzania} & \cellcolor{gray!20}\textbf{Tunisia} \\
    Zimbabwe & & \\
    \hline
    \multicolumn{3}{|c|}{{Americas}} \\
    Argentina & Bolivia & \cellcolor{gray!20}\textbf{Brazil} \\
    Canada & Chile & Colombia \\
    Ecuador & Guatemala & Mexico \\
    \cellcolor{gray!20}\textbf{Paraguay} & Peru & United States \\
    \cellcolor{gray!20}\textbf{Uruguay} & & \\
    \hline
    \multicolumn{3}{|c|}{{Asia}} \\
    Bangladesh & China & India \\
    Iran & \cellcolor{gray!20}\textbf{Iraq} & \cellcolor{gray!20}\textbf{Israel} \\
    Japan & Lebanon & \cellcolor{gray!20}\textbf{Myanmar} \\
    Nepal & North Korea & Pakistan \\
    Saudi Arabia & South Korea & \cellcolor{gray!20}\textbf{Syria} \\
    Turkey & & \\
    \hline
    \multicolumn{3}{|c|}{{Europe}} \\
    Albania & Bulgaria & Denmark \\
    France & Germany & Hungary \\
    Italy & Poland & Romania \\
    Russia & \cellcolor{gray!20}\textbf{Spain} & Sweden \\
    Ukraine & United Kingdom & \\
    \hline
    \multicolumn{3}{|c|}{{Oceania}} \\
    Australia & & \\
    \hline
    \end{tabular}
    \label{wheat_countries}
\end{tabular}
\end{table}

\end{document}